\documentclass[nolinenum]{jfp}

\PassOptionsToPackage{x11names}{xcolor}
\usepackage{jigsaw}
\usepackage{tikz}
\usepackage{stmaryrd}
\usepackage{mathtools}
\usepackage{etoolbox}
\usepackage{idris2}
\usepackage{catchfilebetweentags}
\input{robust-catch}
\usepackage{hyperref}
\usepackage{cleveref}
\usepackage{bytefield}

\usepackage{todonotes}
\setuptodonotes{inline}

\usepackage{listings}

\usepackage{pgfplots}
\pgfplotsset{compat=1.16, width=\textwidth}

\newtoggle{BLIND}

\newcommand{\usestt}{\small}

\newcommand{\agda}{Agda}
\newcommand{\idris}{Idris~2}

\newcommand{\assertTotal}{\IdrisPostulate{assert\KatlaUnderscore{}total}}
\newcommand{\believeMe}{\IdrisPostulate{believe\KatlaUnderscore{}me}}

\newcommand{\hexadesc}[1]{\texttt{\usestt\IdrisType{#1}}}
\newcommand{\hexadata}[1]{\texttt{\usestt\IdrisData{#1}}}
\newcommand{\hexacons}[1]{\texttt{\usestt\IdrisFunction{#1}}}
\newcommand{\hexaoffset}[1]{\texttt{\usestt{\color{lightgray}#1}}}

\newenvironment{hexdump}{\usestt\medskip\ttfamily\obeyspaces\obeylines\noindent}{\medskip}

\newcommand{\suppfile}[1]{File \texttt{#1}}

\usepackage{amsthm}
\newtheorem{remark}{Idris Specifics}

\usepackage{xifthen}
\newcommand{\Pointer}[3]{\ensuremath{#1 \xmapsto{~#2~} #3}}
\newcommand{\Hoare}[4]{\ensuremath{\lbrace~ #1 ~\rbrace~ #2 ~\lbrace~ #3.\, #4 ~\rbrace}}

\togglefalse{BLIND}

\usepackage[T1]{fontenc}
\usepackage{microtype}

\DisableLigatures[<,>]{encoding=T1,family=tt*}

\begin{document}

\journaltitle{JFP}
\cpr{Cambridge University Press}
\doival{10.1017/xxxxx}

\totalpg{\pageref{lastpage01}}
\jnlDoiYr{2022}

\title{Seamless, Correct, and Generic Programming over Serialised Data}

\begin{authgrp}
\author{Guillaume Allais}
\affiliation{University of Strathclyde, G1 1XH\\
        (\email{guillaume.allais@ens-lyon.org})}
\end{authgrp}


\begin{abstract}
In typed functional languages, one can typically only manipulate data
in a type-safe manner if it first has been deserialised into an in-memory
tree internally represented as a graph of nodes-as-structs and subterms-as-pointers.
We demonstrate how we can use Quantitative Type Theory
as implemented in the dependently typed programming language \idris{} to define
a small universe of serialised datatypes, and provide generic programs
allowing users to process values stored contiguously in buffers.
The code manipulating buffer-bound values is extremely similar
to its pure counterpart processing inductive structures thus
allowing a seamless user experience.
Our approach allows implementors to prove the full functional
correctness by construction of the IO functions processing the
data stored in the buffer.
We finally observe how this approach gives us a significant speedup
for functions that do not need to explore the full tree.

This work has been implemented in Idris 2 and fully ported to Agda,
allowing programs written in the two languages to easily exchange
data.
\end{abstract}

\maketitle

\section{Introduction}\label{sec:intro}

In (typed) functional language we are used to manipulating
structured data by pattern-matching on it.
We include an illustrative example below.

\smallskip\noindent
\begin{minipage}{.45\textwidth}
  \begin{center}
    
\newcommand{\mknode}[3]{\draw (#1,#2)  circle (.27cm) node[align=center] {\IdrisData{#3}};}
\newcommand{\mkleaf}[2]{\draw[fill=black] (#1,#2) node[align=center] {} +(-.1cm,-.1cm) rectangle +(.1cm,.1cm);}

\begin{tikzpicture}
\mknode{0}{0}{10}
  \mknode{-1}{-1}{5};
    \mknode{-2}{-2}{1};
      \mkleaf{-2.7}{-3};
      \mkleaf{-1.3}{-3};
  \mkleaf{-.2}{-2}
  \mknode{1}{-1}{20}
    \mkleaf{.2}{-2}
    \mkleaf{1.8}{-2}

\draw [->] (-0.27,0) to [out=180,in=90] (-1,-.73);
  \draw [->] (-1.27,-1) to [out=180,in=90] (-2,-1.73);
    \draw [->] (-2.27,-2) to [out=180,in=90] (-2.7,-2.9);
    \draw [->] (-1.73,-2) to [out=0,in=90] (-1.3,-2.9);
  \draw [->] (-.73,-1) to [out=0,in=90] (-.2,-1.9);
\draw [->] (0.27,0) to [out=0,in=90] (1,-.73);
  \draw [->] (.73,-1) to [out=180,in=90] (.2,-1.9);
  \draw [->] (1.27,-1) to [out=0,in=90] (1.8,-1.9);
\end{tikzpicture}

  \end{center}
\end{minipage}\hfill
\begin{minipage}{.45\textwidth}
  \ExecuteMetaData[Motivating.idr.tex]{motivation}
\end{minipage}
\smallskip

On the left, an example of a binary tree storing bytes in its nodes and
nothing at its leaves.
On the right, a small \idris{} snippet declaring the corresponding
inductive type \IdrisType{Tree} with two constructors \IdrisData{Leaf}
and \IdrisData{Node} and defining a function \IdrisFunction{sum}
adding up all of the nodes' contents.
It proceeds by pattern-matching on its input \IdrisBound{t}:
if the tree is a leaf then we immediately return 0,
otherwise we start by recursively summing up the left and right subtrees, cast the
byte to a natural number and add everything up.
Simply by virtue of being accepted by the typechecker, we know that
this function is covering (all the possible patterns have been handled)
and total (all the recursive calls are performed on smaller trees).

\begin{remark}[Semantic Highligthing]
  All of the code snippets in this paper are semantically
  highlighted using Katla~\citep{MANUAL:repo/github/Katla23}.
  This means that they all have been successfully parsed
  and scope-and-type checked.
  Different colours are used for \IdrisKeyword{keywords},
  \IdrisData{data constructors}, \IdrisFunction{function definitions},
  \IdrisType{types}, \IdrisPostulate{unsafe postulates},
  and \IdrisBound{bound variables}.
\end{remark}

At runtime, the tree will quite probably be represented by
constructors-as-structs and substructures-as-pointers:
each constructor will be a struct with a tag indicating which
constructor is represented and subsequent fields will store
the constructors' arguments.
Each argument will either be a value (e.g. a byte) or a pointer
to either a boxed value or a substructure.
If we were to directly write a function processing a value in this
encoding, proving that a dispatch over a tag is covering, and that
the pointer-chasing is terminating relies on global invariants
tying the encoding to the inductive type.
Crucially, the functional language allows us to ignore all of these
details and program at a higher level of abstraction where we can
benefit from strong static guarantees.

Unfortunately not all data comes structured as inductive values
abstracting over a constructors-as-structs and substructures-as-pointers
runtime representation.
Data that is stored in a file or received over the network is
typically represented in a contiguous binary format with
a high information density.

We include below a textual representation of the above tree using
\texttt{node} and \texttt{leaf} constructors and highlighting the
data in red.

\begin{center}
  \usestt
  \texttt{(node (node (node leaf \IdrisData{1} leaf) \IdrisData{5} leaf) \IdrisData{10} (node leaf \IdrisData{20} leaf))}
\end{center}

This looks almost exactly like the list of bytes we get when using a
naïve serialisation format based on a left-to-right in-order traversal
of this tree.
In the encoding below,
leaves are represented by the byte \texttt{00},
and nodes by the byte \texttt{01}
(each byte is represented by two hexadecimal characters,
we have additionally once again \IdrisData{highlighted} the bytes
corresponding to data stored in the nodes):

\begin{center}
  \usestt
      \texttt{01 $\overbrace{\texttt{01 $\underbrace{\texttt{01 00 \IdrisData{01} 00}}_{\texttt{(node leaf \IdrisData{1} leaf)}}$
    \IdrisData{05} 00}}^{\texttt{(node (node leaf \IdrisData{1} leaf) \IdrisData{5} leaf)}}$
    \IdrisData{0a} 01 00 \IdrisData{14} 00}
\end{center}

The idiomatic way to process such data in a functional language
is to first deserialise it as a value of the inductive type \IdrisType{Tree}
and then call
the tree-processing \IdrisFunction{sum} function we defined above.
If we were using a lower-level language however, we could directly
process the serialised data without the need to fully deserialise it.
Even a naïve port of \IdrisFunction{sum} to C can indeed work
directly over buffers:

{\usestt \begin{lstlisting}
int sumAt (uint8_t buf[], int *ptr) {
  uint8_t tag = buf[*ptr]; (*ptr)++;
  switch (tag) {
    case 0: return 0;
    case 1:
      int m = sumAt(buf, ptr);
      uint8_t b = buf[*ptr]; (*ptr)++;
      int n = sumAt(buf, ptr);
      return (m + (int) b + n);
    default: exit(-1); }}
\end{lstlisting}
}

This function takes a buffer of bytes, and
a pointer currently indicating the start of a tree
and returns the corresponding sum.
We start (line 2) by reading the byte the pointer is referencing and
immediately move the pointer past it.
This is the tag indicating which constructor is at the root of the tree
and so we inspect it (line 3).
If the tag is 0 (line 4), the tree is a leaf and so we return $0$ as the sum.
If the tag is 1 (line 5), then the tree starts with a node and the rest
of the buffer contains
first the left subtree,
then the byte stored in the node,
and finally the right subtree.
We start by summing the left subtree recursively (line 6),
after which the pointer has been moved past its end and is now pointing
at the byte stored in the node.
We can therefore dereference the byte and move the pointer past it (line 7),
compute the sum over the right subtree (line 8),
and finally add up all the components, not forgetting to cast the byte to an int (line 9).
If the tag is anything other than 0 or 1 (line 10) then the buffer does not
contain a valid tree and so we immediately exit with an error code.

As we can readily see, this program
directly performs pointer arithmetic,
explicitly mentions buffer reads,
and relies on undocumented global invariants
such as the structure of the data stored in the buffer,
or the fact that the pointer is being moved along and points directly past
the end of a subtree once \texttt{sumAt} has finished computing
its sum.

Our goal with this work is to completely hide all of these
dangerous aspects
and offer the user the ability to program over serialised data
just as seamlessly and correctly as
if they were processing inductive values.
We will see that
Quantitative Type Theory (QTT)~\citep{DBLP:conf/birthday/McBride16, DBLP:conf/lics/Atkey18}
as implemented in \idris{}~\citep{DBLP:conf/ecoop/Brady21}
empowers us to do just that purely in library code.

\subsection{Seamless Programming over Serialised Data}\label{sec:seamless}

Forgetting about correctness for now, the seamlessness of
our approach can be summed up by the
the following code snippet
in which we compute the sum of the bytes
present in a binary tree stored in a buffer.

\ExecuteMetaData[SaferIndexed.idr.tex]{rsum}

We reserve for later our detailed explanations of the concepts
used in this snippet
(\IdrisType{Pointer.Mu} in \Cref{sec:pointers},
\IdrisFunction{view} in \Cref{sec:dataview}).
For now, it is enough to understand that the function
is an \IdrisType{IO} process
inspecting a buffer that contains a tree stored in serialised format
and computing the same sum as the pure function seen in the previous section.
In both cases, if we uncover a leaf
({\IdrisData{"Leaf"} \IdrisData{\#}} \IdrisKeyword{\KatlaUnderscore{}})
then we return zero,
and if we uncover a node
({\IdrisData{"Node"} \IdrisData{\#}} \IdrisBound{l} \IdrisData{\#} \IdrisBound{b} \IdrisData{\#} \IdrisBound{r})
with
a left branch \IdrisBound{l},
a stored byte \IdrisBound{b},
and a right branch \IdrisBound{r},
then we recursively compute the sums for the left and right subtrees,
cast the byte to a natural number and add everything up.
Crucially, this function and its pure counterpart defined in the previous
section look eerily similar, thanks to the fact that the one operating on
serialised data uses high level combinators and does not explicitly
perform error-prone pointer arithmetic, or low-level buffer reads.

This seamlessness is the first way in which our approach shines.

\subsection{Correct Programming over Serialised Data}

One major difference between the two functions seen above is that
we can easily prove some of the pure function's properties by a structural
induction on its input whereas we
cannot prove anything about the \IdrisType{IO} process without first
explicitly postulating the \IdrisType{IO} monad's properties.
We will see that we can instead refine that second definition to obtain
a correct-by-construction version of
\IdrisFunction{sum}, with almost exactly the same code.

\ExecuteMetaData[SaferIndexed.idr.tex]{csum}

In the above snippet, we can see that the \IdrisType{Pointer.Mu} is indexed
by a phantom parameter: a runtime irrelevant \IdrisBound{t} which has type
(\IdrisType{Data.Mu} \IdrisFunction{Tree}).
And so the return type can mention the result of the pure computation
(\IdrisFunction{Data.sum} \IdrisBound{t}).
\IdrisType{Singleton} is, as its name suggests, a singleton type
(cf. \Cref{sec:view})
i.e. it is a wrapper around a natural number that is proven to be equal to
the one computed by the pure \IdrisFunction{sum} function.
And so we can transfer any property proven on the pure \IdrisFunction{sum}
to the one operating on values residing in buffers.

The implementation itself only differs in that we had to use idiom
brackets~\citep{DBLP:journals/jfp/McbrideP08}, something we will explain
in \Cref{sec:datasingleton}.

In other words, our approach also allows us to, \emph{at the same time},
define and prove the functional correctness of the \IdrisType{IO}
procedures processing trees stored in serialised format in a buffer.
This means in particular that any intermediate computation can rely
on the fact that the recursive calls are already known to be correct.

This intrinsic correctness is our second main contribution.

\subsection{Generic Programming over Serialised Data}

Last but not least,
as Altenkirch and McBride~\citeyearpar{DBLP:conf/ifip2-1/AltenkirchM02}
demonstrated:
``With dependently (sic) types, generic programming is just programming:
it is not necessary to write a new compiler each time a useful
universe presents itself.''

In this paper we carve out a universe of inductive types that can be
uniformly serialised and obtain all of our results by generic programming.
In practice this means that we are not limited to the type of binary trees
with bytes stored in the nodes we used in the examples above.
We will for instance be able to implement
a generic and correct-by-construction
definition of \IdrisFunction{fold} operating on data stored in a buffer
whose type declaration can be seen below
(we will explain the pure generic fold in \Cref{sec:genericfoldinductive}
and define its counterpart operating on buffers in \Cref{sec:bufferfold}).

\ExecuteMetaData[SaferIndexed.idr.tex]{foldtype}

This data-genericity is our third contribution.

\subsection{Plan}

In summary, we are going to define a library for the
seamless,
correct,
and generic
manipulation of algebraic types in serialised format.

\Cref{sec:desc} introduces the language of descriptions capturing the
subset of inductively defined types that our work can handle.
It differs slightly from usual presentations in that it ensures the
types can be serialised and tracks crucial invariants towards that goal.
\Cref{sec:trees} gives a standard meaning to these data descriptions
as strictly positive endofunctors whose fixpoints give us the expected
inductive types.
We will use this standard meaning in the specification layer of our work.
\Cref{sec:hexdump} explores the serialisation format we have picked
for these trees: a depth-first, left-to-right infix traversal of the
trees, with additional information stored to allow for the direct access
of any subtree.
\Cref{sec:pointers} defines the type of pointers to trees stored in a
buffer and shows how we can use such pointers to write the corresponding
tree to a file.
\Cref{sec:view} introduces the terminology of \emph{views} and
\emph{singleton} types that is crucial to the art of programming
in a correct-by-construction manner.
\Cref{sec:poking} defines IO primitives that operate on serialised
trees stored in an underlying buffer.
They encapsulate all the unsafe low-level operations and offer a
high-level interface that allows users to implement correct-by-construction
procedures.
\Cref{sec:serialising} defines a set of serialisation combinators that
allows users to implement correct-by-construction procedures writing
values into a buffer.
\Cref{sec:timing} discusses some preliminary performance results for
the library.

\section{Our Universe of Descriptions}\label{sec:desc}

We first need to pin down the domain of our discourse.
To talk generically about an entire class of datatypes
without needing to modify the host language
we have decided to perform a universe
construction~\citep{DBLP:journals/njc/BenkeDJ03, DBLP:phd/ethos/Morris07, DBLP:conf/icfp/LohM11}.
That is to say that we are going to introduce an inductive type
defining a set of codes together
with an interpretation of these codes as bona fide
host-language types.
We will then be able to program generically over the universe of
datatypes by performing induction on the type of
codes~\citep{DBLP:conf/tphol/PfeiferR99}.

The universe we define is in the tradition of
a sums-of-products vision of inductive types~\citep{DBLP:conf/popl/JanssonJ97}.
In our setting, constructors are essentially arbitrarily nested tuples of
values of type unit,
bytes,
and recursive substructures.
A datatype is given by listing a choice of constructors.

\subsection{Interlude: Serialisation Formats}\label{sec:interludeserialisation}

Before giving a precise definition of the descriptions, let us think
a little bit about the features we would like to see in our serialisation
format and see how this informs our design.
We reproduce below the naïve encoding of a binary tree with bytes stored
in its nodes we gave in the introduction. It has both strengths and
weaknesses.

\begin{center}
  \usestt
      \texttt{01 $\overbrace{\texttt{01 $\underbrace{\texttt{01 00 \IdrisData{01} 00}}_{\texttt{(node leaf \IdrisData{1} leaf)}}$
    \IdrisData{05} 00}}^{\texttt{(node (node leaf \IdrisData{1} leaf) \IdrisData{5} leaf)}}$
    \IdrisData{0a} 01 00 \IdrisData{14} 00}
\end{center}

One of its strength is the flattened encoding of nodes:
the left subtree, byte stored, and right subtree are all stored
contiguously after the tag \texttt{01} announcing that a \texttt{node} is present.
This approach can be generalised to any type with a unique constructor:
there is no need to store a byte corresponding to the constructor for the
unit type, or for the constructor to a pair type. These unambiguous
constructors can be reinserted during the type-directed decoding phase.
In practice this means that our encoding will not retain the nesting
structure described by a type. Concretely, $((a, b), c)$ and $(a, (b, c))$
will have the same representations: the encoding of $a$, followed by the
encoding of $b$ and last that of $c$.
Similarly, values of type $()$ will be eluded entirely: values of type
$((), a, ())$ and $a$ will have the same serialised representation, the
encoding of the value of type $a$.

The major weakness of the naïve format we presented above is the
inability to process a node's right subtree without having first
processed its left subtree.
This is due to the fact that the subtree's size is not statically
known and that the serialisation format does not store an offset
for it.
Consequently, if we wanted to fetch the rightmost byte stored
in the tree, we would have to traverse the entire encoding. That
represents an exponential slow-down compared to what we could achieve
with a different encoding allowing for random access.

Based on these observations, we envision a serialisation format
that will permit random access to all of the arguments of a node.
Such a format will have to store an offset for all of a node's
arguments whose size is not statically known and which we may want
to jump past i.e. all of them except a constructor's very last argument.

\subsection{Descriptions}

This leads us to the inductive family \IdrisType{Desc} declared below
as the type of constructor descriptions.

\ExecuteMetaData[Serialised/Desc.idr.tex]{desctype}

\begin{remark}[Separate Declaration and Definition]
  \idris{} lets us declare a datatype or function first and only
  define it later. This is convenient for blocks of mutually-defined
  types and functions but can also be used, like here, to focus on
  the type signature first.
\end{remark}

This family has three indices corresponding to three crucial
invariants being tracked.
First, an index telling us whether the current description
is being used in the \IdrisBound{rightmost} branch of the overall
constructor description.
Second, the \IdrisBound{static}ally known size of the described data
in the number of bytes it occupies.
Third, the number of \IdrisBound{offsets} that need to be stored to
compensate for subterms not having a statically known size.
The reader should think of \IdrisBound{rightmost} as an `input' index
(the context in which the family is used tells it whether it is in
a rightmost branch)
whereas \IdrisBound{static} and \IdrisBound{offsets} are `output' indices
(the family's own constructors each determine their respective sizes).

Next we define the family proper by giving its four constructors.

\ExecuteMetaData[Serialised/Desc.idr.tex]{desc}

Each constructor can be used anywhere in a description so their index
tracking whether we are in the rightmost branch can be an arbitrary
boolean \IdrisBound{r}.

\IdrisData{None} is the description of values of type unit. The static
size of these values is zero as no data is stored in a value of type unit.
Similarly, they do not require an offset to be stored as we statically
know their size.

\IdrisData{Byte} is the description of bytes.
Their static size is precisely one byte, and they do not require an
offset to be stored either.

\IdrisData{Prod} gives us the ability to pair two descriptions together.
As explained earlier pairs have a unique constructor and so their encoding
will not consume any additional bytes.
Consequently their static size and number of offsets are the respective
sums of the static sizes and numbers of offsets of each subdescription.
The description of the left element of the pair will never be in the
rightmost branch of the overall constructors description and so its
index is \IdrisData{False} while the description of the right element
of the pair is in the rightmost branch precisely whenever the whole pair
is; hence the propagation of the \IdrisBound{r} arbitrary value from the
return index into the description of the right component.

Last but not least, \IdrisData{Rec} is a position for a subtree.
We cannot know its size in bytes statically and so we decide to store
an offset unless we are in the rightmost branch of the overall description.
Indeed, there are no additional constructor arguments behind the rightmost
one and so we have no reason to skip past the subterm. Consequently we
do not bother recording an offset for it.

\subsection{Constructors}

We represent a constructor as a record packing together
a name for the constructor,
the description of its arguments (which is, by virtue of
being used at the toplevel, in rightmost position),
and the values of the \IdrisFunction{static} and
\IdrisFunction{offsets} invariants.

\ExecuteMetaData[Serialised/Desc.idr.tex]{constructor}

\begin{remark}[Implicit Record Fields]
  Record fields whose type declaration is surrounded by
  curly braces are implicit: they do not need to be explicitly
  mentioned when constructing the record, or when pattern-matching
  on it.
\end{remark}

Here the two invariants \IdrisFunction{static} and
\IdrisFunction{offsets} are stored as implicit fields
because as `output' indices of the \IdrisType{Desc} family
their value is easily reconstructed using
unification~\citep{DBLP:conf/tlca/AbelP11}.
Note that we used \IdrisData{(::)} as the name of the
constructor for records of type \IdrisType{Constructor}.
This allows us to define constructors by forming an
expression reminiscent of Haskell's type declarations:
\IdrisBound{name} \IdrisData{::} \IdrisBound{type}.
Returning to our running example, this gives us the following encodings for
leaves that do not store anything
and nodes that contain a left branch, a byte, and a right branch.

\noindent
\begin{minipage}[t]{.38\textwidth}
  \ExecuteMetaData[Serialised/Desc.idr.tex]{treeleaf}
\end{minipage}\hfill
\begin{minipage}[t]{.58\textwidth}
  \ExecuteMetaData[Serialised/Desc.idr.tex]{treenode}
\end{minipage}

\subsection{Datatypes}

A datatype description \IdrisType{Data}
is given by a vector (also known as
a length-indexed list) named \IdrisFunction{constructors}
and containing constructor descriptions.
For convenience, we store the number \IdrisFunction{consNumber}
of constructors separately so that it does not need to be
recomputed every time it is needed.
We additionally insist that we have a proof \IdrisFunction{fitsInBits8}
that the datatype has less than 255 constructors which will allow us
to safely store the corresponding tag in a single byte.
This is enforced using \IdrisType{So}, a type
family ensuring the boolean check it is indexed by has succeeded.

\ExecuteMetaData[Serialised/Desc.idr.tex]{data}\label{sec:datadescriptions}

\begin{remark}[Auto Implicit Record Fields]
  Implicit record fields whose type declaration is preceded by the
  \IdrisKeyword{auto} keyword are auto implicit: just like implicit
  fields they do not need to be explicitly mentioned when constructing
  the record, or when pattern-matching on it.
  Additionally, these values are not constructed by unification but
  via a builtin type-directed proof search.
\end{remark}

We can then encode our running example as a simple \IdrisType{Data}
declaration: a binary tree whose node stores bytes is described by the choice
of either a \IdrisFunction{Leaf} or \IdrisFunction{Node}, as defined above.
The proof \IdrisFunction{fitsInBits8} is constructed automatically:
the boolean test (\IdrisData{2} \IdrisFunction{<=} \IdrisData{255})
has trivially computed to \IdrisData{True}.

\ExecuteMetaData[Serialised/Desc.idr.tex]{treedesc}\label{fig:treedesc}

Now that we have a language that allows us to give a description of our
inductive types, we are going to give these descriptions a meaning as trees.

\section{Meaning as Trees}\label{sec:trees}

We now see descriptions as functors and, correspondingly,
datatypes as the initial objects of the associated functor-algebras.
This is a standard construction derived from Malcolm's
work~\citeyearpar{DBLP:journals/scp/Malcolm90},
itself building on Hagino's categorically-inspired
definition of a lambda calculus
with a generic notion of datatypes~\citep{DBLP:conf/ctcs/Hagino87}.

Intuitively, the meaning of a description is the shape of one ``layer'' of
tree where the subtrees' meaning is left abstract. We include below two
sketches showing what the respective meanings of the \IdrisFunction{Node}
and \IdrisFunction{Leaf} descriptions look like.
We use jigsaw pieces to figure ports in which subtrees can be inserted.

\noindent
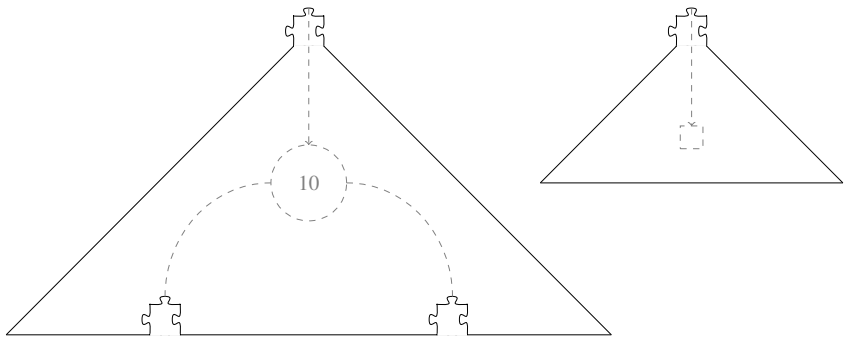
\begin{center}
\scalebox{.8}{
\colorlet{bgcolor}{white}
\tikzset{
    overdraw/.style={preaction={draw,bgcolor,line width=#1}},
    overdraw/.default=1pt
}

\newcommand{\interface}[2]{
  \begin{scope}[xshift=-7.2+(5*5.4*#1), yshift=-7.2+(5*5.4*#2), scale=.5]
    \piece{0}{1}{-1}{-1};
    \draw[overdraw, color=white] (.015, 0) to (.99, 0);
  \end{scope}
}

\newcommand{\mknode}[3]{\draw[color=gray, dashed] (#1,#2)  circle (2.5cm) node[align=center] {#3};}
\newcommand{\mkleaf}[2]{\draw[color=gray, dashed] (#1,#2) node[align=center] {} +(-.75cm,-.75cm) rectangle +(.75cm,.75cm);}

\begin{tikzpicture}

  \interface{0}{0}
  \interface{-2.5}{-5}
  \interface{2.5}{-5}

  \begin{scope}[scale=.25]
    \draw (-1, -1) to (-20, -20);
    \draw (1, -1) to (20, -20);

    \draw (-20, -20) to (-10.5, -20);
    \draw (-8.5, -20) to (8.5, -20);
    \draw (10.5, -20) to (20, -20);

    \draw [->, dashed, color=gray] (0, 1.5) to (0, -7.5);
    \mknode{0}{-10}{10}
    \draw [dashed, color=gray] (-2.5, -10) to [out=180, in=90] (-9.5, -17.5);
    \draw [dashed, color=gray] (2.5, -10) to [out=0, in=90] (9.5, -17.5);
  \end{scope}

  \begin{scope}[xshift=180]
  \interface{0}{0};

  \begin{scope}[scale=.25]
    \draw (-1, -1) to (-10, -10);
    \draw (1, -1) to (10, -10);

    \draw (-10, -10) to (10, -10);

    \draw [->, dashed, color=gray] (0, 1.5) to (0, -6.25);
    \mkleaf{0}{-7}
  \end{scope}
  \end{scope}
\end{tikzpicture}
}
\end{center}

In our work these trees will be used primarily to allow users to
give a precise specification of the IO procedures they actually want
to write in order to process values stored in buffers.
We expect these inductive trees and the associated generic programs
consuming them to be mostly used at the \IdrisKeyword{0}
modality i.e. to be erased during compilation.

\subsection{Descs as Functors}

We are going to define the \IdrisFunction{Meaning} of descriptions
as strictly positive endofunctors on \IdrisType{Type} by
induction on said descriptions.
In its type, all of \IdrisBound{r}, \IdrisBound{s},
and \IdrisBound{o} are implicitly universally quantified,
a feature of \idris{} we explain below.

\ExecuteMetaData[Serialised/Desc.idr.tex]{meaningtype}

\begin{remark}[Implicit Prenex Polymorphism]\label{rmk:prenexpoly}
  Lowercase names that are seemingly unbound are automatically
  quantified over in a prenex manner reminiscent of other functional
  languages like OCaml or Haskell.
  These variables are bound at quantity 0, meaning that they will
  be automatically erased during compilation.
\end{remark}

This function will interpret every \IdrisType{Desc} constructor
as its obvious meaning as a type, using the parameter to give a
meaning to \IdrisData{Rec} positions.
In particular, \IdrisData{None} and \IdrisData{Prod} will
respectively be interpreted by a unit type (\IdrisType{True})
and a product type (\IdrisType{Tuple}) defined below.
We do not use the unit and product from the standard
library purely to offer better syntactic sugar to
our users.

\noindent
\begin{minipage}[t]{.4\textwidth}
  \ExecuteMetaData[Lib.idr.tex]{true}
\end{minipage}\hfill
\begin{minipage}[t]{.5\textwidth}
  \ExecuteMetaData[Lib.idr.tex]{pair}
\end{minipage}

\begin{remark}[Lack of Eta-rules for Records]\label{rmk:etarecords}
  \IdrisType{True} and \IdrisType{Tuple} are both defined
  as \IdrisKeyword{record}s.
  However in \idris{}, this is currently only syntactic sugar
  to declare a datatype together with a projection for each
  of the record's fields.
  Conversion checking does not incorporate eta rules for records
  and so we have to manually state, prove, and deploy eta rules
  whenever necessary.
  We include the two lemmas below as we will need them later.

  \ExecuteMetaData[Lib.idr.tex]{etaTrue}\label{fig:etatrue}
  \ExecuteMetaData[Lib.idr.tex]{etaTuple}\label{fig:etatuple}
\end{remark}

Now that we have these two type constructors, we can explicitly
define \IdrisFunction{Meaning}, the function giving
us the action on objects of the \IdrisType{Desc}-encoded
endofunctors.

\ExecuteMetaData[Serialised/Desc.idr.tex]{meaningfun}

Both \IdrisData{None} and \IdrisData{Byte} are interpreted by constant
functors (respectively the one returning the unit type \IdrisType{True},
and the one returning the type of bytes).
(\IdrisData{Prod} \IdrisBound{d} \IdrisBound{e})
is interpreted as a tuple grouping the respective interpretations
of \IdrisBound{d} and \IdrisBound{e}.
Finally \IdrisData{Rec} is the identity functor.


Now that we have the action of descriptions on types,
let us see their action on morphisms: provided a
function from \IdrisBound{a} to \IdrisBound{b}, we can
build one from the \IdrisFunction{Meaning} of \IdrisBound{d}
at type \IdrisBound{a} to its meaning at type \IdrisBound{b}.
We once again proceed by induction on the description.

\ExecuteMetaData[Serialised/Desc.idr.tex]{fmap}\label{def:fmap}

All cases but the one for \IdrisData{Rec} are structural.
We leave out the proofs verifying that these definitions
respect the functor laws up to pointwise equality.
They are included in the supplementary material
(\suppfile{Data.Serialisable.Data}).

\subsection{Data as Trees}

Given a datatype description \IdrisBound{cs}, our first goal is
to define what it means to pick a constructor.
The \IdrisType{Index} record is a thin wrapper around a
(\IdrisType{Fin} (\IdrisFunction{consNumber} \IdrisBound{cs}) i.e.
a finite natural number known to be smaller than the number of
constructors the \IdrisBound{cs} type provides.

\ExecuteMetaData[Serialised/Desc.idr.tex]{index}

We use this type rather than \IdrisType{Fin} directly because it
plays well with inference.
In the following code snippet, implementing a function returning
the description corresponding to a given index,
we use this to our advantage:
the \IdrisBound{cs} argument can be left implicit because it already
shows up in the type of the \IdrisType{Index} and can thus be
reconstructed by unification~\citep{DBLP:conf/tlca/AbelP11}.

\ExecuteMetaData[Serialised/Desc.idr.tex]{description}

The \IdrisFunction{index} function is provided by the standard
library: given a position of type (\IdrisType{Fin} \IdrisBound{n})
and a vector of size \IdrisBound{n}, it returns the value located
at that position.
Note that we seem to define \IdrisFunction{description}
in terms of itself, in a manner that would create an
infinite loop.
But the occurrence on the right-hand side is actually
referring to the projection out of the
\IdrisType{Constructor} record.
This is possible thanks to Idris' type-directed disambiguation.

\begin{remark}[Type-directed Disambiguation]
  If multiple definitions in scope have the same name,
  Idris performs type-directed disambiguation to pick
  the only one that would work in that context.
\end{remark}

This type of indices also allows us to provide users with
syntactic sugar enabling them to use the constructors' names
directly rather than confusing numeric indices.
The following function runs a semi-decision procedure
\IdrisFunction{isConstructor} (whose implementation is
not given here) at the type level
in order to turn any raw string \IdrisBound{str}
into the corresponding \IdrisType{Index}.

\ExecuteMetaData[Serialised/Desc.idr.tex]{fromString}

\begin{remark}[Ad-hoc Polymorphisms for Literals]
  String, numeric, and floating point literals are respectively
  desugared using \IdrisFunction{fromString}, \IdrisFunction{fromInteger}, and
  \IdrisFunction{fromDouble}.
  Combined with the type-directed disambiguation (see above)
  of overloaded symbols, this allows users to compute potentially complex
  data from literals in a type-directed manner.
\end{remark}

In this instance, this allows us to use string literals as
proxies for constructor names.
If the string literal stands for a valid name then
\IdrisFunction{isConstructor} will
return a valid \IdrisType{Index} and the compiler's proof
search mechanism will be able to
\IdrisKeyword{auto}matically fill-in the implicit proof.
If the name is not valid then \idris{} will not
find the index and will raise a compile time error.
We include a successful example on the left and a failing test
on the right hand side where the compiler is not able to find
a proof of (\IdrisFunction{isJust} \IdrisKeyword{(}
\IdrisFunction{isConstructor} \IdrisData{"Cons"}
\IdrisFunction{Tree}\IdrisKeyword{)}) because it is
simply not the case that \IdrisData{"Cons"} is the
name of a \IdrisFunction{Tree} constructor.

\noindent
\begin{minipage}[t]{0.35\textwidth}
  \ExecuteMetaData[Serialised/Desc.idr.tex]{indexleaf}
\end{minipage}\hfill
\begin{minipage}[t]{0.52\textwidth}
\ExecuteMetaData[Serialised/Desc.idr.tex]{notindexcons}
\end{minipage}

\begin{remark}[Failing Blocks]
  A \IdrisKeyword{failing} block is a way to document (and
  enforce) that some code leads to an error.
  Such blocks are only accepted if their body parses but
  leads to an error during elaboration.
\end{remark}

Once equipped with the ability to pick constructors, we can define
the type of algebras for the functor described by a \IdrisType{Data}
description. For each possible constructor, we demand an algebra for
the functor corresponding to the meaning of the  constructor's description.

\ExecuteMetaData[Serialised/Desc.idr.tex]{alg}

We can then introduce the fixpoint of data descriptions as the initial
algebra, defined as the following inductive type.

\ExecuteMetaData[Serialised/Desc.idr.tex]{mu}

\begin{remark}[Escape hatches]
  \idris{} provides some escape hatches to use when the author
  knows a usage is safe but the compiler is not able to
  determine that it is.

  A function call of the form (\assertTotal{} $e$) will
  circumvent the termination and positivity checkers in
  the expression $e$.
  This is only safe if the function is actually terminating
  or the type strictly positive.

  A function call of the form (\believeMe{} $e$) will be
  usable at any type. This is only safe if $e$'s actual type
  and the type it is being used at have the same runtime
  representation.
\end{remark}

Note that here we are forced to use \assertTotal{} to convince \idris{}
to accept the definition.
Indeed, unlike Agda, \idris{} does not (yet!) track whether a function's
arguments are used in a strictly positive manner.
Consequently the positivity checker
is unable to see that \IdrisFunction{Meaning} uses its second
argument in a strictly positive manner,
and so that \IdrisFunction{Alg} also is,
and that this is therefore a legal definition.

Now that we can build trees as fixpoints of the
meaning of descriptions, we can define convenient aliases for
the \IdrisFunction{Tree} constructors.
Note that the leftmost \IdrisData{(\#)} use in each definition corresponds
to the \IdrisType{Mu} constructor while later ones are \IdrisType{Tuple}
constructors.
\idris{}'s type-directed disambiguation of constructors allows us to use
this uniform notation for all of these pairing notions.

\noindent
\begin{minipage}[t]{.3\textwidth}
  \ExecuteMetaData[Serialised/Desc.idr.tex]{leaf}
\end{minipage}\hfill
\begin{minipage}[t]{.65\textwidth}
  \ExecuteMetaData[Serialised/Desc.idr.tex]{node}
\end{minipage}

This enables us to define our running example as an inductive value
of type (\IdrisType{Mu} \IdrisFunction{Tree}):

\ExecuteMetaData[Serialised/Desc.idr.tex]{longexample}

\subsection{Generic Fold}\label{sec:genericfoldinductive}

\IdrisType{Mu} gives us the initial fixpoint for these algebras i.e.
given any other algebra over a type \IdrisBound{a}, from a term of
type (\IdrisType{Mu} \IdrisBound{cs}), we can compute an \IdrisBound{a}.
We define the generic \IdrisFunction{fold} function over inductive values
as follows:

\ExecuteMetaData[Serialised/Desc.idr.tex]{fold}

We first match on the term's top constructor, use \IdrisFunction{fmap}
(defined in \Cref{def:fmap})
to recursively apply the fold to all the node's subterms and finally
apply the algebra to the result.
Here we only use \assertTotal{} because \idris{} does not see that
\IdrisFunction{fmap} only applies its argument to strict subterms.
This limitation could easily be bypassed by mutually defining
an inlined and specialised version of
(\IdrisFunction{fmap} \IdrisKeyword{\KatlaUnderscore} (\IdrisFunction{fold} \IdrisBound{alg}))
as we demonstrate in \Cref{appendix:safefold}.
In an ideal type theory these supercompilation steps, whose sole
purpose is to satisfy the totality checker, would be automatically
performed by the compiler~\citep{MANUAL:phd/dublin/Mendel12}.


Further generic programming can yield other useful programs e.g. a
generic proof that tree equality is decidable,
a generic definition of zippers~\citep{DBLP:conf/icfp/LohM11},
or a tail-recursive version of fold~\citep{DBLP:conf/icfp/CortinasS18}.

\section{Serialised Representation}\label{sec:hexdump}

Before we can give a meaning to descriptions as pointers into a
buffer we need to decide on a serialisation format.
The format we have opted for is split in two parts: a header, followed
by the actual representation of the tree as a contiguous block of bytes.
The header contains an encoding of the description of the type of the
tree stored in the rest of the buffer.
It can be used when loading a file to check that a user's claim about
the presence of a serialised tree of a given type is correct.

For instance, the following binary snippet is a hex dump of a file
containing the serialised representation of a binary tree belonging to
the type we have been using as our running example.
The raw data is semantically highlighted:
8-bytes-long\footnote{This implicitly means that we assume
that our terms occupy less than $2^{64}$ bytes,
a very reasonable assumption in our opinion.}
\hexaoffset{offsets} in little endian format,
a \hexadesc{type} description of the stored data,
some \hexacons{nodes} of the tree
and the \hexadata{data} stored in the nodes.

\begin{center}

\begin{hexdump}
87654321\hphantom{:} 00 11 22 33 44 55 66 77 88 99 AA BB CC DD EE FF
00000000: \hexaoffset{07 00 00 00 00 00 00 00} \hexadesc{02 00 02 03 02 01 03} \hexacons{01}
00000010: \hexaoffset{17 00 00 00 00 00 00 00} \hexacons{01} \hexaoffset{0c 00 00 00 00 00 00}
00000020: \hexaoffset{00} \hexacons{01} \hexaoffset{01 00 00 00 00 00 00 00} \hexacons{00} \hexadata{01} \hexacons{00} \hexadata{05} \hexacons{00} \hexadata{0a}
00000030: \hexacons{01} \hexaoffset{01 00 00 00 00 00 00 00} \hexacons{00} \hexadata{14} \hexacons{00} \hphantom{00} \hphantom{00} \hphantom{00} \hphantom{00}
\end{hexdump}

\end{center}

More specifically, this block is the encoding of the \IdrisFunction{example}
given in the previous section and,
accepting that the tree's representation starts at byte \texttt{FF}, and
knowing that a \IdrisFunction{leaf} is represented here by \hexacons{00}
and a \IdrisFunction{node} is represented by \hexacons{01}
readers can check
(ignoring the offsets for now)
that the data is stored in a depth-first, left-to-right traversal of the tree
i.e. we get exactly the bit pattern we saw in the naïve encoding
presented in \Cref{sec:intro}:

\begin{center}
  \usestt
      \texttt{\IdrisFunction{01} \hexaoffset{$\cdots$} $\overbrace{\texttt{\IdrisFunction{01} \hexaoffset{$\cdots$}  $\underbrace{\texttt{\IdrisFunction{01} \hexaoffset{$\cdots$}  \IdrisFunction{00} \IdrisData{01} \IdrisFunction{00}}}_{\texttt{(\IdrisFunction{\footnotesize{node}} \IdrisFunction{\footnotesize{leaf}} \IdrisData{1} \IdrisFunction{\footnotesize{leaf}})}}$
    \IdrisData{05} \IdrisFunction{00}}}^{\texttt{(\IdrisFunction{\footnotesize{node}} (\IdrisFunction{\footnotesize{node}} \IdrisFunction{\footnotesize{leaf}} \IdrisData{1} \IdrisFunction{\footnotesize{leaf}}) \IdrisData{5} \IdrisFunction{\footnotesize{leaf}})}}$
    \IdrisData{0a} \IdrisFunction{01} \hexaoffset{$\cdots$}  \IdrisFunction{00} \IdrisData{14} \IdrisFunction{00}}
\end{center}

Let us now look at the format more closely.

\subsection{Header}

In our example, the header is as follows:
\begin{hexdump}
\hexaoffset{07 00 00 00 00 00 00 00} \hexadesc{02 00 02 03 02 01 03}
\end{hexdump}

The header consists of an 8-bytes long offset stored in little endian
followed by a binary representation of the
\IdrisType{Data} description of the value stored in the buffer.
The offset allows us to jump past the description in case we do
not care to inspect it.

This description can be useful in a big project where different
components produce and consume such serialised values:
if we change the format in one place but forget to update
it in another, we want the program to gracefully
fail to load the file using an unexpected format.
We detail in \Cref{sec:limitation-robust} how dependent
type providers~\citep{DBLP:conf/icfp/Christiansen13}
can help structure a software project
to prevent such issues at compile time.

The encoding of a data description starts with a byte giving us the number
of constructors\footnote{
Remember that we enforce in the definition of the \IdrisType{Data} record
type in \Cref{sec:datadescriptions}
that our descriptions cannot have more than 255 constructors. So using
a single byte here is safe.},
followed by these constructors' respective descriptions
serialised one after the other.
\IdrisData{None} is represented by \hexadesc{00},
\IdrisData{Byte} is represented by \hexadesc{01},
(\IdrisData{Prod} \IdrisBound{d} \IdrisBound{e}) is represented by
\hexadesc{02} followed by the representation of \IdrisBound{d} and then that of \IdrisBound{e},
and \IdrisData{Rec} is represented by \hexadesc{03}.

Looking once more at the header in the running example,
the \IdrisType{Data} description is indeed 7 bytes long like the offset states.
The \IdrisType{Data} description starts with \hexadesc{02}
meaning that the type has two constructors.
The first one is \hexadesc{00} i.e. \IdrisData{None}
(this is the encoding of the type of \IdrisFunction{Leaf}),
and the second one is \hexadesc{02 03 02 01 03} i.e.
\IdrisKeyword{(}\IdrisData{Prod} \IdrisData{Rec}
\IdrisKeyword{(}\IdrisData{Prod} \IdrisData{Byte} \IdrisData{Rec}\IdrisKeyword{))}
(that is to say the encoding of the type of \IdrisFunction{Node}).
According to the header, this file does contain a \IdrisFunction{Tree}.

\subsection{Tree Serialisation}\label{sec:tree-serialisation}

The design of this format was guided by a few simple principles
detailed in \Cref{sec:interludeserialisation}:
\begin{enumerate}
  \item the format should be contiguous (no pointer indirections)
  \item values of type unit should occupy no space
  \item the nesting of pairs in a description should have no impact on the layout
  \item the format should support the direct access to any of a node's subtrees
\end{enumerate}
This last criterion allows us to skip past subtrees
that we do not need to process thus ensuring an exponential
speedup compared to formats forcing an in-order traversal of
a node's subtrees.
To this end each node needs to store an offset measuring the size of the
subtrees that are to the left of any relevant information.

If a given tag is associated to a description of type
(\IdrisType{Desc} \IdrisData{True} \IdrisBound{s} \IdrisBound{o})
then the representation in memory of the associated node will look something
like the following.
On the first line we have a description of the data layout and on the
second line we have the offset of various positions in the block with
respect to the tag's address.

\label{fig:data-layout}
\begin{center}
\begin{bytefield}[bitwidth=6mm, bitheight=7mm]{4}
  \bitbox{2}[bgcolor=Chartreuse4!40]{tag}
  & \bitbox{4}[bgcolor=lightgray!30]{$o$ offsets}
  & \bitbox[ltb]{2}{tree$_1$}
  & \bitbox[tb]{1}{$\cdots$}
  & \bitbox[tb]{2}[bgcolor=IndianRed1!40]{byte$_1$}
  & \bitbox[tb]{1}{$\cdots$}
  & \bitbox[tb]{2}{tree$_k$}
  & \bitbox[tb]{1}{$\cdots$}
  & \bitbox[tb]{2}[bgcolor=IndianRed1!40]{byte$_s$}
  & \bitbox[tbr]{2}{tree$_{o+1}$} \\
  \bitbox[l]{1}{\small$0$}
  & \bitbox[]{1}{}
  & \bitbox[l]{1}{\small$1$}
  & \bitbox[]{3}{}
  & \bitbox[l]{3}{\small$1+8\times{}o$}
  & \bitbox[]{6}{}
  & \bitbox[l]{5}{\small$8\times{}o+s+\Sigma_{i=1}^{o}o_i$}
\end{bytefield}

\end{center}

For the data layout,
we start with the tag
then we have $o$ offsets,
and finally we have a block contiguously storing an interleaving of
subtrees and $s$ bytes
dictated by the description.
In this example the rightmost value in the description is a subtree and
so even though we have $o$ offsets, we actually have $(o+1)$ subtrees stored.

The offsets of the tag with respect to its own address is $0$.
The tag occupies one byte and so the offset of the block of offsets is $1$.
Each offset occupies 8 bytes and so the constructor's arguments
are stored at offset $(1+8\times{}o)$.
Finally each value's offset can be computed by adding up
the offset of the start of the block of constructor arguments,
the offsets corresponding to all of the subtrees that come before it,
and the number of bytes stored before it;
in the case of the last byte that gives $1+8\times{}o + \Sigma_{i=1}^{o}o_i + s-1$
hence the formula included in the diagram.

Going back to our running example, this translates to the following
respective data layouts and offsets for a leaf and a node.

\begin{center}
  \begin{bytefield}[bitwidth=6mm, bitheight=7mm]{4}
  \bitbox[]{2}{Leaf} \\
  \bitbox{2}[bgcolor=Chartreuse4!40]{00} \\
  \bitbox[l]{1}{\small$0$}
\end{bytefield}\qquad
\begin{bytefield}[bitwidth=.05\linewidth, bitheight=7mm]{4}
  \bitbox[]{2}{Node} \\
  \bitbox{2}[bgcolor=Chartreuse4!40]{01}
  & \bitbox{4}[bgcolor=lightgray!30]{offset}
  & \bitbox{4}{left subtree}
  & \bitbox{2}[bgcolor=IndianRed1!40]{byte}
  & \bitbox{4}{right subtree} \\
  \bitbox[l]{1}{\small$0$}
  & \bitbox[]{1}{}
  & \bitbox[l]{1}{\small$1$}
  & \bitbox[]{3}{}
  & \bitbox[l]{1}{\small$9$}
  & \bitbox[]{3}{}
  & \bitbox[l]{2}{\small$9+o_1$\hfill}
  & \bitbox[l]{2}{\small$10+o_1$}
\end{bytefield}

\end{center}

We now have a good understanding of the serialisation format we
are going to use to represent our inductive trees.
The next step is to define what it means to have a pointer to a
tree residing in a buffer.

\section{Interlude: Logics for Imperative Programs}

Let us first look back at some logical frameworks built
to explicitly talk about memory locations and their contents
in order to prove the properties of imperative programs.
Their design and properties will then inform our choices.

\subsection{Hoare Logic}

Hoare logic~\citeyearpar{DBLP:journals/cacm/Hoare69}  is a framework in
which we can state and prove the properties of imperative programs.
Its central notion is that of Hoare triples. They are statements of
the form
\[ \Hoare{P}{e}{v}{Q} \]
declaring that under the precondition
$P$, and binding the result of evaluating the expression $e$ as $v$,
we can prove that $Q$ holds.

The most basic of predicates to express knowledge about the memory
state is a `points to' assertion (\Pointer{\ell}{}{w}) stating that
the label $\ell$ points to a memory location containing the word $w$.
\label{sec:deref}
This can be used for instance to specify the behaviour of a primitive
dereferencing operator: if $\ell$ is known to point to a value $w$ then evaluating
the program (\texttt{deref}($\ell$)) will return a value $v$ equal to $w$.
\[ \Hoare{\Pointer{\ell}{}{w}}{\texttt{deref}(\ell)}{v}{v = w} \]

Provided that we only have read-only operations, evaluating a program cannot
invalidate assumptions about the contents of the memory and so we enjoy a
form of weakening: for any proposition $R$, if we can specify the behaviour
of a program $e$ using the precondition $P$ and the postcondition $Q$,
then the same program will also abide by the specification using the
precondition ($P \wedge R$) and the postcondition ($Q \wedge R$).
\[ \Hoare{P}{e}{v}{Q} \quad \Longrightarrow \quad \Hoare{P \wedge R}{e}{v}{Q \wedge R} \]
In particular, the validity of (\Pointer{\ell}{}{w}) can be propagated even
after having dereferenced it.

\newcommand{\letin}[3]{\texttt{let}\,#1\,=\,#2~\texttt{in}~#3}

Using further structural rules, we can combine the axioms to compositionally
prove statements about more complex programs. We present below a structural
rule for a \texttt{let} construct: provided that
executing $e$ returns a value $v$ satisfying $Q$ under precondition $P$ and that
for all $x$, executing $e^{\prime}$ returns a value $v$ satisfying $R$ under
precondition $Q(x)$
then executing ($\letin{x}{e}{e^{\prime}}$) will return a value $v$ satisfying $R$
under precondition $P$.

\begin{gather*}
  \Hoare{P}{e}{v}{Q} \wedge \forall x. \Hoare{Q(x)}{e^{\prime}}{v}{R} \\
  \Longrightarrow \Hoare{P}{\letin{x}{e}{e^{\prime}}}{v}{R}
\end{gather*}

We can for instance prove that we can chase pointers:
if $\ell$ points to a value $\ell^{\prime}$ itself corresponding
to a label pointing to $w$, nested calls to \texttt{deref}
will return a value equal to $w$:
\[ \Hoare
     {\Pointer{\ell}{}{\mathit{\ell^{\prime}}} \wedge \Pointer{\ell^{\prime}}{}{\mathit{w}}}
     {\letin{x}{\texttt{deref}(\ell)}{\texttt{deref}(x)}}{v}{v = w}
\]

Once we have these building blocks, it becomes interesting to introduce
some abstractions. As we explained above, (\Pointer{\ell}{}{w}) states that
the memory location called $\ell$ contains the word $w$.
The programs we are interested in do not however manipulate isolated words,
they talk about full blown inductive trees.
The key idea is to use this base `points to' assertion as a building
block to compositionally give a meaning to richly typed pointers.
We will write (\Pointer{\ell}{\mathit{ty}}{v}) for the assumption that
the label $\ell$ points to a value $v$ of type $\mathit{ty}$.
We can naturally define (\Pointer{\ell}{\text{\IdrisType{Bits8}}}{w}) as
simply (\Pointer{\ell}{}{w}).
We can then decide that, assuming that we know the size $s_a$
of values of type $a$, (\Pointer{\ell}{(a,~ b)}{(v_1, v_2)})
means that $\ell$ points to two contiguous values $v_1$ and $v_2$.
In other words it is an alias for
($\Pointer{\ell}{a}{v_1} \wedge \Pointer{\ell+s_a}{b}{v_2}$).

These richly typed pointers are the kind of pointers we are going to
define in the next section.
But before, let us have a look at how things get more complicated
once we are given access to primitives that can destructively update
the contents of a memory location.

\subsection{Separation Logic}

It is tempting to add a new primitive \texttt{assign} that takes a
label $\ell$ and a word $w$ and updates the memory cell so that
(\Pointer{\ell}{}{w}) now holds true. Its specification would be
\[ \Hoare{\Pointer{\ell}{}{\_}}{\texttt{assign}\,(\ell,\,w)}{\_}{\Pointer{\ell}{}{w}} \]
However with such a primitive things that were true before a program
was run may have been invalidated in the process.
Consequently the weakening principle given above does not hold anymore.
Otherwise the following reasoning step would be valid and the memory cell
corresponding to the label $\ell$ would need to contain both $0$ and $1$.
\begin{gather*}
  \Hoare{\Pointer{\ell}{}{0}}{\texttt{assign}\,(\ell,\,1)}{\_}{\Pointer{\ell}{}{1}}
  \\\quad \centernot\Longrightarrow \quad
   \Hoare{\Pointer{\ell}{}{0} \wedge \Pointer{\ell}{}{0}}{\texttt{assign}\,(\ell,\,1)}{\_}{\Pointer{\ell}{}{1} \wedge \Pointer{\ell}{}{0}}
\end{gather*}

This makes Hoare logic highly non-compositional in the presence of
destructive updates: having proven a program
fragment's specification using predicates mentioning the memory locations
it interacts with, this result cannot be reused when proving the specification
of a larger program involving additional memory locations.

The state-of-the-art solution is to move away from Hoare logic and use
separation logic~\citep{DBLP:conf/lics/Reynolds02,DBLP:journals/cacm/OHearn19,DBLP:books/hal/Chargueraud23,MANUAL:book/sfoundations/Chargueraud23}
whose core idea is to have a \emph{separating} conjunction: ($P \ast Q$)
states that both $P$ and $Q$ hold true but that they are talking about
\emph{separate} slices of the memory.
This allows for the safe inclusion of a similar weakening principle,
the \emph{frame rule} which states that for an arbitrary predicate $R$
if $e$ can be specified using precondition $P$ and postcondition $Q$
then it also can for precondition ($P \ast R$) and postcondition ($Q \ast R$)
precisely because $R$ is only talking about memory locations not impacted by $e$.

As Rouvoet~\citeyearpar{DBLP:phd/basesearch/Rouvoet21}
demonstrated, it is possible to embed separation logic in
type theory but, even with best efforts, it remains a somewhat
heavy process. Luckily for us, we do not need to.

Our goal with this work is to process data serialised in buffers
without first deserialising it as an in-memory tree.
To do so we only need to be able to read from the buffer to
analyse the structure of the inductive data stored in it.
Although we may \emph{want} to have destructive updates in future
work, for this specific task we do not actually \emph{need} them.
Correspondingly, we will happily stick to a Hoare-style logic and
its really powerful weakening principle.
As a matter of fact, we will go even further and entirely bypass the
usual explicit embedding of such logics in the host language.
We will use \idris{}'s abstraction facilities to introduce richly typed
`points to' predicates as first class values that can just be passed
around, and trusted primitives that can take these pointers and reveal
the shape of the values they are pointing to.

\section{Meaning as Pointers Into a Buffer}\label{sec:pointers}

For reasons that will become apparent in \Cref{sec:bufferfold}
when we start programming over serialised data in a correct-by-construction
manner, our types of `pointers' will be parameterised not only
by the description of the type of the data stored but also by a
runtime-irrelevant inductive value of that type.

Note that we will \emph{not} export the constructors to the various
`pointer' types defined in the following section.
Consequently, the only way for a user to get their hands on such a
pointer is to use the library functions we provide.
By only providing invariant-respecting functions, we can ensure
that the assumptions encoded in the phantom parameters are never
violated.

\subsection{Tracking Buffer Positions}

We start with the definition of \IdrisType{Pointer.Mu},
the counterpart to \IdrisType{Data.Mu} for serialised values.

\ExecuteMetaData[SaferIndexed.idr.tex]{pointermu}

A tree sitting in a buffer is represented
by a record packing the buffer, the position at which the tree's
root node is stored, and the size of the tree.
The record is indexed by \IdrisBound{cs} a \IdrisType{Data} description
and \IdrisBound{t} the tree of type (\IdrisType{Data.Mu} \IdrisBound{cs})
which is represented by the buffer's content.
Neither are mentioned in the types of the record's fields, making them
\emph{phantom types}~\citep{DBLP:conf/dsl/LeijenM99}.
A term of type (\IdrisType{Pointer.Mu} \IdrisBound{cs} \IdrisBound{t})
plays a double role in our library: it acts both as a label $\ell$ in
the runtime relevant layer, and a proof that said label points to a
buffer-bound value equal to $t$
(i.e. \Pointer{\ell}{\text{\IdrisType{Data.Mu} \IdrisBound{cs}}}{t}).


The pointer counterpart to a \IdrisFunction{Meaning} stores
additional information.

\ExecuteMetaData[SaferIndexed.idr.tex]{elem}

For a description of type (\IdrisType{Desc} \IdrisBound{r} \IdrisBound{s} \IdrisBound{o})
on top of the buffer, the position at which the root of the meaning resides,
and the size of the layer we additionally have a vector of \IdrisBound{o} offsets
corresponding to the sizes of the subtrees we may want to skip past.
This will allow us to efficiently access any constructor argument we want.

\subsection{Writing a Tree to a File}\label{sec:writetofile}

Once we have a pointer to a tree \IdrisBound{t} of type \IdrisBound{cs}
(\IdrisType{Pointer.Mu} \IdrisBound{cs} \IdrisBound{t} in the type below)
in a buffer, we can easily write it to a file be it for safekeeping
or sending over the network.

\ExecuteMetaData[SaferIndexed.idr.tex]{writeToFile}

\begin{remark}[Forall Quantifier]
  The \IdrisKeyword{forall} quantifier is sugar for an implicit
  binder at quantity \IdrisKeyword{0}.
  It can be useful to introduce variables that cannot be automatically
  bound in a prenex manner because they have a type dependency over an
  explicitly bound argument.
\end{remark}

We first start by reading the size of the header stored in the buffer.
This allows us to compute both the \IdrisBound{start} of the data block
as well as the size of the buffer (\IdrisBound{bufSize}) that will
contain the header followed by the tree we want to write to a file.
We then check whether the position of the pointer is exactly the beginning
of the data block.
If it is then we are pointing to the whole tree and the current buffer can
be written to a file as is.
Otherwise we are pointing to a subtree and need to separate it from its
surrounding context first.
To do so we allocate a new buffer of the right size and use the
standard library's \IdrisFunction{copyData} primitive to copy the raw bytes
corresponding to the header first, and the tree of interest second.
We can then write the buffer we have picked to a file and happily succeed.

\subsection{Reading a Tree from a File}

We can also go in the other direction: using the data description
\IdrisBound{cs}, we can load the content of the file located at
\IdrisBound{fp} as a pointer to the root of the tree of type
(\IdrisType{Mu} \IdrisBound{cs}) we claim is contained in it.

\ExecuteMetaData[SaferIndexed.idr.tex]{readFromFile}

This function takes a default argument \IdrisBound{safe} controlling
whether we should attempt to check that the file starts with a header
containing a type descripton matching the one passed as an argument
by the caller.

\begin{remark}[Default Arguments]
  An implicit argument can be assigned a \IdrisKeyword{default} value.
  It will take this value unless explicitly overwritten by the caller.
\end{remark}

First, we create a buffer from the content of the file.
We then read the offset giving us the size of the header.
If we want to be \IdrisBound{safe} we then read the type description
contained in the header using \IdrisFunction{getData} (not shown here)
and check it for equality against the one we were passed using \IdrisFunction{eqData}
(not shown here).
If the check fails, we emit an error and fail.
If the check succeeds, we postulate the existence of a runtime irrelevant
tree meant to represent the file's content and put together a pointer
for that asbtract tree.
This postulate cements the user's claim about the file's content;
naturally if the file is not in fact contain a serialised representation
of a tree this can lead to fatal errors later on when attempting to
inspect the buffer's content.
We return a pair of the runtime irrelevant tree and a pointer to it
thus ensuring users cannot directly attempt to match on the inductive
tree; they will have to use the combinators we are about to define to
inspect it by reading into the buffer.

Note that, in order to save space in the paper, we never checked whether
the buffer reads we performed in both \IdrisFunction{writeToFile} and
\IdrisFunction{readFromFile} were within bounds.
A released version of the library would naturally need to include such
checks.

Now that we have pointers,
and use files to read and write the trees they are standing for,
we are only missing the ability to look at the content they are pointing to.
But first we need to introduce some basic tools
to be able to talk precisely about this stored content.

\section{Interlude: Views and Singletons}\label{sec:view}

The precise indexing of pointers by a runtime-irrelevant copy of the value
they are pointing to means that inspecting the buffer's content should
not only return runtime information but also refine the index to reflect
that information at the type-level.
As a consequence, the buffer-inspecting functions we are going to define
will be views.

\subsection{Views}

A view
in the sense of Wadler~\citep{DBLP:conf/popl/Wadler87},
and subsequently refined by McBride and McKinna~\citep{DBLP:journals/jfp/McBrideM04}
for a type $T$ is a type family $V$ indexed by $T$ together
with a function which maps values $t$ of type $T$ to values of type
$V\,t$.
By inspecting the $V\,t$ values we can learn something about the
$t$ input.
The prototypical example is perhaps the `snoc` (`cons' backwards) view
of right-nested lists as if they were left-nested.
We present the \IdrisType{Snoc} family below.

\ExecuteMetaData[Snoc.idr.tex]{Snoc}

By matching on a value of type
(\IdrisType{Snoc} \IdrisBound{xs}) we get to learn
either that \IdrisBound{xs} is empty (\IdrisData{Lin}, nil backwards)
or that it has an initial segment \IdrisBound{init} and a last element
\IdrisBound{last} (\IdrisBound{init} \IdrisData{:<} \IdrisBound{last}).
Crucially (\IdrisBound{init} \IdrisFunction{++} \IdrisData{[}\IdrisBound{last}\IdrisData{]})
is not a valid pattern because it mentions a stuck function call
but (\IdrisBound{init} \IdrisData{:<} \IdrisBound{last}) is as
it is only made up of constructors and binding positions.
And so by having a function that computes the (\IdrisType{Snoc} \IdrisBound{xs})
view of any list \IdrisBound{xs}, we are able to pretend as if we were
actually able to match on ``patterns'' of the form
(\IdrisBound{init} \IdrisFunction{++} \IdrisData{[}\IdrisBound{last}\IdrisData{]}).
The function \IdrisFunction{unsnoc} demonstrates that we can always
view a \IdrisType{List} in such a \IdrisType{Snoc}-manner.

\ExecuteMetaData[Snoc.idr.tex]{unsnoc}

\begin{remark}[With-Abstraction]
  The \IdrisKeyword{with} construct allows programmers to locally
  define an anonymous auxiliary function taking an extra argument
  compared to its parent.
  By writing (\IdrisKeyword{with} \IdrisKeyword{(}$e$\IdrisKeyword{)})
  we introduce such an auxiliary function an immediately apply it to $e$.
  The nested clauses that immediately follow each take an extra pattern
  which matches over the possible values of $e$.
  If the left-hand side of the auxiliary function is the same as that of
  its parents bar the pattern for the newly added argument, we can use the
  elision notation (\IdrisKeyword{\_} \IdrisKeyword{|}) to avoid having to
  repeat ourselves.

  In other words the following definition of \IdrisFunction{f} using a
  \IdrisKeyword{with} construct with the elision notation

  \ExecuteMetaData[With.idr.tex]{fsugar}

  \noindent is equivalent to the following desugared version where
  the auxiliary function \IdrisFunction{fAux}
  taking an extra argument has been lifted
  to the toplevel.

  \ExecuteMetaData[With.idr.tex]{fdesugar}

\end{remark}

In the code snippet for \IdrisFunction{unsnoc} we performed a recursive call on
(\IdrisFunction{unsnoc} \IdrisBound{xs}) and distinguished
two cases: when the view returns the empty snoclist \IdrisData{[<]}
and when it returns an \IdrisBound{init}ial segment together with the
\IdrisBound{last} element.
Because we are using a view, matching on these constructors actually
refined the shape of the parent clause's argument \IdrisBound{xs}.
We do not need to spell out its exact shape in each branch because
we were careful to only introduce \IdrisBound{xs} as a name alias
using an as-pattern while letting the actual pattern be a catch-all
pattern (\IdrisBound{xs}\IdrisKeyword{@}\IdrisKeyword{\_}).
This is a common trick to make working with views as lightweight as
possible.

Here we defined \IdrisType{Snoc} as an inductive family but it can
sometimes be convenient to define the family recursively instead,
in which case the \IdrisType{Singleton} inductive family can
help us connect runtime values to their
runtime-irrelevant type-level counterparts.

\subsection{The Singleton Type}\label{sec:datasingleton}

The \IdrisType{Singleton} family has a single constructor
which takes an argument \IdrisBound{x} of type \IdrisBound{a},
its return type is indexed precisely by this \IdrisBound{x}.

\ExecuteMetaData[Data/Singleton.idr.tex]{singleton}

More concretely this means that a value of type
(\IdrisType{Singleton} $t$) has to be a runtime relevant
copy of the term $t$.
Note that \idris{} performs an optimisation similar to Haskell's
\texttt{newtype} unwrapping: every datatype that has a single
non-recursive constructor with only one non-erased argument
is unwrapped during compilation.
This means that at runtime the
\IdrisType{Singleton} / \IdrisData{MkSingleton} indirections
will have disappeared.

We can define some convenient combinators to manipulate
singletons.
We reuse the naming conventions typical of applicative
functors which will allow us to rely on \idris{}'s automatic
desugaring of \emph{idiom brackets}~\citep{DBLP:journals/jfp/McbrideP08}
into expressions using these combinators.

First \IdrisFunction{pure} is a simple alias for \IdrisData{MkSingleton},
it turns a runtime-relevant value \IdrisBound{x} into a singleton for
this value.

\ExecuteMetaData[Data/Singleton.idr.tex]{pure}

Next, we can `map' a function under a \IdrisType{Singleton} layer: given
a pure function \IdrisBound{f} and a runtime copy of \IdrisBound{t} we
can get a runtime copy of (\IdrisBound{f} \IdrisBound{t}).

\ExecuteMetaData[Data/Singleton.idr.tex]{fmap}

Finally, we can apply a runtime copy of a function \IdrisBound{f}
to a runtime copy of an argument \IdrisBound{t}
to get a runtime copy of the result (\IdrisBound{f} \IdrisBound{t}).

\ExecuteMetaData[Data/Singleton.idr.tex]{ap}

As we mentioned earlier, \idris{} automatically desugars idiom brackets
using these combinators.

\begin{remark}[Idiom Brackets as Sugar]
  Idiom brackets let us use the standard whitespace-based application
  typical of pure functions to performs computations in an \emph{Applicative}
  contexts.
  In other words, during elaboration an atomic expression
  \IdrisKeyword{[|} \IdrisBound{x} \IdrisKeyword{|]}
  will be desugared to (\IdrisFunction{pure} \IdrisBound{x})
  while a compound expression
  \IdrisKeyword{[|} \IdrisBound{f} \IdrisBound{t1} $\cdots$ \IdrisBound{tn} \IdrisKeyword{|]}
  will become (\IdrisBound{f} \IdrisFunction{<\$>} \IdrisBound{t1} \IdrisFunction{<*>} $\cdots$ \IdrisFunction{<*>} \IdrisBound{tn}).
\end{remark}

This built-in handling of idiom brackets lets us apply
\IdrisType{Singleton}-wrapped values almost as seamlessly
as pure values.

We are now equipped with the appropriate notions and definitions to
look at a buffer's content.

\section{Inspecting a Buffer's Content}\label{sec:poking}

Let us now describe the combinators allowing our users to
take apart the values they have a pointer for.
These functions will read bytes in the buffer and reflect
the observations thus made at the type level by refining
the pointer's indices.

There will be two separate tiers of definitions: the most
basic building blocks (\IdrisFunction{poke} and \IdrisFunction{out})
will be a trusted core of primitives implemented using
escape hatches.
This is inevitable given that we are reflecting the content
of buffer reads at the type level.
We will clearly specify their behaviour by explaining what
benign Hoare-style axioms they correspond to.

We will then show how we can use these low-level trusted
primitives to define higher level combinators
(\IdrisFunction{layer} and \IdrisFunction{view}).
Crucially these definitions will not need to use further
escape hatches: provided that the trusted core is correct,
then so will they.
This will culminate in the implementation of a generic
correct-by-construction version of \IdrisFunction{fold}
operating over trees stored in a buffer (cf. \Cref{sec:bufferfold}).

\subsection{Poking the Buffer}\label{sec:poke}

Our most basic operation consists in poking the buffer to gain
access to the head constructor of the underlying layer of
\IdrisFunction{Data.Meaning} we have a pointer to.
This operation is description-directed and so its result (called
\IdrisFunction{Poke}) is defined by case analysis on the description
associated to the pointer.

Concretely, the type of the function is as follows: provided a pointer for
a description \IdrisBound{d}, subtrees of type \IdrisBound{cs} and an
associated meaning \IdrisBound{t} of type
(\IdrisFunction{Meaning} \IdrisBound{cs} \IdrisBound{t})
we return an \IdrisType{IO} process computing the one step
unfolding of the meaning.

\ExecuteMetaData[SaferIndexed.idr.tex]{pokefun}

As we explained, \IdrisFunction{Poke} is defined by case-analysis
on the description.
However, in order to keep the notations user-friendly, we
are forced by Idris' lack of eta-rules (cf. \cref{rmk:etarecords})
to mutually define
an inductive family \IdrisType{Poke'} with interesting return
indices.
It will allow users to, by matching on \IdrisType{Poke'}
constructors, automatically refine the associated meaning
present at the type level into a term with a head constructor.
This will ensure that functions defined by pattern-matching
can reduce in types based on observations made at the term
level.

\ExecuteMetaData[SaferIndexed.idr.tex]{pokedatafun}

Poking a buffer containing a \IdrisData{Byte} will yield a
runtime-relevant copy of the type-level byte we have for
reference, hence the use of \IdrisType{Singleton}.
This corresponds to adding the following Hoare-style axiom
for \IdrisFunction{poke}. Remembering that
(\IdrisFunction{Meaning} \IdrisData{Byte} (\IdrisType{Mu} \IdrisBound{cs}))
computes to \IdrisType{Bits8} and so that
(\Pointer{\ell}{\text{\IdrisFunction{Meaning} \IdrisData{Byte} (\IdrisType{Mu} \IdrisBound{cs})}}{w})
is essentially (\Pointer{\ell}{}{w}), we
note that this Hoare-style axiom looks eerily
similar to the axiom for \texttt{deref} we gave in
\Cref{sec:deref}):
\[ \Hoare
     {\Pointer{\ell}{\text{\IdrisFunction{Meaning} \IdrisData{Byte} (\IdrisType{Mu} \IdrisBound{cs})}}{w}}
     {\text{\IdrisFunction{poke}}~\ell}
     {v}{v = w}
\]
If the description is \IdrisData{Rec} this means
we have a substructure. In this case we simply demand
a pointer to it. This amounts to adding the following
axiom:
\[ \Hoare
     {\Pointer{\ell}{\text{\IdrisFunction{Meaning} \IdrisData{Rec} (\IdrisType{Mu} \IdrisBound{cs})}}{t}}
     {\text{\IdrisFunction{poke}}~\ell}
     {\ell^{\prime}}{\Pointer{\ell^{\prime}}{\text{\IdrisType{Mu} \IdrisBound{cs}}}{t}}
\]

Last but not least, if we are accessing a value of a
record type (either \IdrisData{None} or a \IdrisData{Prod} of two descriptions)
then we describe the resulting observation using the \IdrisType{Poke'} family.

\ExecuteMetaData[SaferIndexed.idr.tex]{pokedatadata}

\IdrisType{Poke'} has two constructors corresponding to the two
descriptions it covers.
If the description of the buffer's content is \IdrisData{None}
then we do not expect to get a value back, only the knowledge
that the type-level meaning is \IdrisData{I}. This corresponds
to adding the following axiom.
\[ \Hoare
     {\Pointer{\ell}{\text{\IdrisFunction{Meaning} \IdrisData{None} (\IdrisType{Mu} \IdrisBound{cs})}}{t}}
     {\text{\IdrisFunction{poke}}~\ell}
     {\_}{t = \text{\IdrisData{I}}}
\]
If the description is (\IdrisData{Prod} \IdrisBound{d} \IdrisBound{e})
then we demand to learn that the type-level term is \IdrisData{(\#)}-headed
with two substructures $t_1$ and $t_2$ and we expect
\IdrisFunction{poke} to give us a pointer to each of these substructures.
This corresponds to the following axiom.
\begin{gather*}
  \Hoare
     {\Pointer{\ell}{\text{\IdrisFunction{Meaning} (\IdrisData{Prod} \IdrisBound{d} \IdrisBound{e}) (\IdrisType{Mu} \IdrisBound{cs})}}{t}}
     {\\\text{\IdrisFunction{poke}}~\ell\\}
     {(\ell_1,~\ell_2)}
     {\exists t_1.~ \exists t_2.~
       t = (t_1 ~\text{\IdrisData{\#}}~ t_2)
       \wedge \Pointer{\ell_1}{\text{\IdrisFunction{Meaning} \IdrisBound{d} (\IdrisType{Mu} \IdrisBound{cs})}}{t_1}
       \wedge \Pointer{\ell_2}{\text{\IdrisFunction{Meaning} \IdrisBound{e} (\IdrisType{Mu} \IdrisBound{cs})}}{t_2}}
\end{gather*}

As we mentioned earlier, the \idris{} implementation of the
\IdrisFunction{poke} function
is necessarily using escape hatches as we are essentially
giving a computational content to the axioms listed above.
We proceed by case analysis on the description.
Let us go through each case one-by-one.

\ExecuteMetaData[SaferIndexed.idr.tex]{pokefunNone}

If the description is \IdrisData{None} we do not need to fetch any
information from the buffer but we do need to deploy the eta rule
for \IdrisType{True} (cf. \Cref{fig:etatrue} for the definition
of \IdrisFunction{etaTrue})
in order to be able to use the \IdrisType{Poke'}
constructor \IdrisData{I}.

\ExecuteMetaData[SaferIndexed.idr.tex]{pokefunByte}

If the description is \IdrisData{Byte} then we read a byte at the
determined position. The only way we can connect this value we just
read to the runtime irrelevant type index is to use the unsafe combinator
\IdrisPostulate{unsafeMkSingleton} to manufacture a value of type
(\IdrisType{Singleton} \IdrisBound{t}) instead of the value of type
(\IdrisType{Singleton} \IdrisBound{bs})
we would expect from wrapping \IdrisBound{bs} in the \IdrisData{MkSingleton} constructor.
As we explained earlier, this amounts to realising the
Hoare-style axiom specifying the act of dereferencing a pointer.

\ExecuteMetaData[SaferIndexed.idr.tex]{pokefunProd}
If the description is the product of two sub-descriptions then we
want to compute the \IdrisType{Pointer.Meaning} corresponding to
each of them.
We do so by following the serialisation format we detailed in
\Cref{sec:hexdump}.
We start by splitting the vector of offsets to distribute them between
the left and right subtrees.

We can readily build the pointer for the \IdrisBound{left} subdescription:
it takes the left offsets, the buffer, and has the same starting position
as the whole description of the product as the submeanings are stored one after the other.
Its size (called \IdrisBound{sizel}) is the sum of the space reserved
by all of the left offsets (\IdrisFunction{sum} \IdrisBound{subl})
as well as the static size occupied by the rest of the content
(\IdrisBound{sl}).

We then compute the starting position of the right subdescription: we need to
move past the whole of the left subdescription, that is to say that the starting
position is the sum of the starting position for the whole product and \IdrisBound{sizel}.
The size of the right subdescription is then easily computed by subtracting
\IdrisBound{sizel} from the overall \IdrisBound{size} of the paired subdescriptions.

We can finally use the lemma \IdrisFunction{etaTuple}
(defined in \Cref{fig:etatuple}) saying that a tuple
is equal to the pairing of its respective projections
in order to turn \IdrisBound{t} into
(\IdrisFunction{fst} \IdrisBound{t} \IdrisData{\#} \IdrisFunction{snd} \IdrisBound{t})
which lets us use the \IdrisType{Poke'} constructor \IdrisData{(\#)} to return our
pair of pointers.

Although we did not need to use escape hatches here, the implementation
is still part of the trusted core in that we are computing offsets in
(we claim!) accordance with the serialisation format.

\ExecuteMetaData[SaferIndexed.idr.tex]{pokefunRec}

Lastly, when we reach a \IdrisData{Rec} description, we can discard the
vector of offsets and return a \IdrisType{Pointer.Mu} with the same buffer,
starting position and size as our input pointer.

\subsection{Extracting One Layer}

By repeatedly poking the buffer, we can unfold a full layer of term.
This operation is not part of the trusted core: provided that
\IdrisFunction{poke} is correct then it will automatically be
correct-by-construction.
The result type is once again defined by induction on the description.
It is essentially identical to the definition of
\IdrisFunction{Poke} except for the \IdrisData{Prod} case:
instead of being content with a pointer for each of the
subdescriptions, we demand a \IdrisFunction{Layer} for them too.

\ExecuteMetaData[SaferIndexed.idr.tex]{layerdata}

This function can easily be implemented by induction on the description
and repeatedly calling \IdrisFunction{poke} to expose the values one by
one.
We call \IdrisFunction{poke} and use the \IdrisType{IO} monad's bind
operator (\IdrisFunction{>>=}) to pass the result to \IdrisFunction{go},
the auxiliary function recursively going under record constructors to
perform further poking.

\ExecuteMetaData[SaferIndexed.idr.tex]{layerfun}

If the description is \IdrisData{None}, \IdrisData{Byte}, or
\IdrisData{Rec} then poking the buffer was already enough to
reveal the full layer and we can simply return the result.
If we have a (\IdrisData{Prod} \IdrisBound{d} \IdrisBound{e})
then poking the buffer revealed that the term is \IdrisData{(\#)}-headed
and handed us a pointer to each of its components.
We call \IdrisFunction{layer} recursively on each of these pointers and
use idiom brackets to combine the \IdrisType{IO}-wrapped results
using \IdrisType{Layer'}'s \IdrisData{(\#)} constructor.

We can readily use this function to inspect meanings
we have a pointer to. In the following artificial example,
looking at the goal \texttt{?hole}, we learn that
\IdrisBound{v} has the shape {\IdrisKeyword{(}\IdrisBound{t} \IdrisData{\#} \IdrisKeyword{(}\IdrisData{I} \IdrisData{\#} \IdrisBound{u}\IdrisKeyword{))}}
where \IdrisBound{t} is a tree of type (\IdrisType{Mu} \IdrisBound{cs})
and \IdrisBound{u} is a byte, and that
\IdrisBound{q} is a pointer to \IdrisBound{t}
and \IdrisBound{w} has type (\IdrisType{Singleton} \IdrisBound{u}).
\ExecuteMetaData[SaferIndexed.idr.tex]{layerexample}
We have effectively managed to take \IdrisBound{v} apart in a
type-directed manner and get a handle on its subterms.
In that sense we have just defined a view for meanings stored in buffers!

\subsection{Exposing a Tree's Top Node}\label{sec:out}

Now that we can deserialise an entire layer of \IdrisFunction{Meaning},
the only thing we are missing to be able to generically manipulate trees
is the ability to expose the top node of a tree stored at a
\IdrisType{Pointer.Mu} position.
This will require the addition of a new function to the trusted core:
the function \IdrisFunction{out}.
Its type states that given a pointer
to a tree \IdrisBound{t} of type \IdrisBound{cs} we can get an
\IdrisType{IO} process revealing the top node of \IdrisBound{t}.

\ExecuteMetaData[SaferIndexed.idr.tex]{outfun}

The \IdrisType{Out} family formally describes what revealing the
top node means:
obtaining an \IdrisType{Index} named \IdrisBound{k},
and a \IdrisType{Pointer.Meaning} to the meaning
\IdrisBound{t} of the description associated to \IdrisBound{k}.
The family's index (\IdrisBound{k} \IdrisData{\#} \IdrisBound{t})
(where the overloaded \IdrisData{(\#)} is here the
\IdrisType{Data.Mu} constructor)
ensures that the structure of the runtime irrelevant
tree is adequately described by this index and meaning.

\ExecuteMetaData[SaferIndexed.idr.tex]{outdata}

This amounts to introducing the following Hoare-style axiom for
\IdrisFunction{out}:
\[
\Hoare
    {\Pointer{\ell}{\text{\IdrisType{Mu} \IdrisBound{cs}}}{t}}
    {\text{\IdrisFunction{out}}\,\ell}
    {(k,\, \ell_1)}
    {\exists t_1.~ t = (k \,\text{\IdrisData{\#}}\, t_1)
      \wedge \Pointer{\ell_1}{\text{\IdrisFunction{Meaning} \IdrisBound{cs$_k$} (\IdrisType{Mu} \IdrisBound{cs})}}{t_1}}
\]

The function \IdrisFunction{out} being part of the trusted
core means that its implementation will once again need to
use escape hatches to reconcile the buffer's observed content
with the type-level indices.
Let us recall the data layout detailed in \Cref{fig:data-layout}:
\begin{center}
\begin{bytefield}[bitwidth=6mm, bitheight=7mm]{4}
  \bitbox{2}[bgcolor=Chartreuse4!40]{tag}
  & \bitbox{4}[bgcolor=lightgray!30]{$o$ offsets}
  & \bitbox[ltb]{2}{tree$_1$}
  & \bitbox[tb]{1}{$\cdots$}
  & \bitbox[tb]{2}[bgcolor=IndianRed1!40]{byte$_1$}
  & \bitbox[tb]{1}{$\cdots$}
  & \bitbox[tb]{2}{tree$_k$}
  & \bitbox[tb]{1}{$\cdots$}
  & \bitbox[tb]{2}[bgcolor=IndianRed1!40]{byte$_s$}
  & \bitbox[tbr]{2}{tree$_{o+1}$} \\
  \bitbox[l]{1}{\small$0$}
  & \bitbox[]{1}{}
  & \bitbox[l]{1}{\small$1$}
  & \bitbox[]{3}{}
  & \bitbox[l]{3}{\small$1+8\times{}o$}
  & \bitbox[]{6}{}
  & \bitbox[l]{5}{\small$8\times{}o+s+\Sigma_{i=1}^{o}o_i$}
\end{bytefield}

\end{center}
Operationally, \IdrisFunction{out}
will amount to inspecting the tag used by the node,
deserialising the offsets stored immediately after it,
and forming a pointer to the start of the meaning block.
As a first step, let us get our hands on the index of the head constructor.
\ExecuteMetaData[SaferIndexed.idr.tex]{getIndex}
We obtain a byte by calling \IdrisFunction{getBits8}, cast it to a
natural number and then make sure that it is in the range
$[\text{\IdrisData{0}} \cdots \text{\IdrisFunction{consNumber} \IdrisBound{cs}}[$ using
\IdrisFunction{natToFin}. If the check fails then we return
a hard fail: the buffer contains an invalid representation
and so the precondition that we only have pointers to valid
values was violated.

We can now describe the \IdrisFunction{out} function's implementation.
\ExecuteMetaData[SaferIndexed.idr.tex]{outfunbody}
We start by reading the index \IdrisBound{k}
corresponding to the constructor choice.
We then use the unsafe \IdrisPostulate{unfoldAs} postulate to step the
type-level \IdrisBound{t} to something of the form
(\IdrisBound{k} \IdrisData{\#} \IdrisBound{val}).
We then conclude using the \IdrisFunction{getConstructor} function
(defined later) to gather the required offsets and put together
the pointer to the meaning of the description associated to \IdrisBound{k}.

\ExecuteMetaData[Serialised/Desc.idr.tex]{unfoldAs}

The declaration of \IdrisPostulate{unfoldAs} is marked as runtime
irrelevant because it cannot possibly be implemented
(\IdrisBound{t} is runtime irrelevant and so cannot be inspected)
and so its output should not be relied upon in runtime-relevant
computations.
Its type states that there exists a \IdrisFunction{Meaning} called
\IdrisBound{val} such that \IdrisBound{t} is equal to
(\IdrisBound{k} \IdrisData{\#} \IdrisBound{val}).
This is of course untrue in general: we cannot take an arbitrary
\IdrisBound{t} and declare that it is \IdrisBound{k}-headed.
This use-case is however justified in that it reflects at the
type-level the observation we made by reading the buffer.

Now that we know the head constructor we want to deserialise and that
we have the ability to step the runtime irrelevant tree to match the
actual content of the buffer, we can use \IdrisFunction{getConstructor}
to build such a value.
Given a pointer to a tree (\IdrisBound{k} \IdrisData{\#} \IdrisBound{t}),
it will read enough information from the buffer to assemble a pointer to
the meaning \IdrisBound{t}.

\ExecuteMetaData[SaferIndexed.idr.tex]{getConstructor}

To get a pointer to the meaning \IdrisBound{t},
we start by getting the vector of offsets stored
immediately after the tag. We then compute the size of the remaining
\IdrisFunction{Meaning} description: it is the size of the overall tree,
minus $1$ (for the tag) and $8$ times the number of offsets (because
each offset is stored as an 8 bytes number).
We can then use the record constructor \IdrisData{MkMeaning} to pack
together the vector of offsets, the buffer, the position past the offsets
and the size we just computed.

\ExecuteMetaData[SaferIndexed.idr.tex]{getOffsets}

The implementation of \IdrisType{getOffsets} is straightforward: given
a continuation that expect \IdrisBound{n} offsets as well as the
position past the last of these offsets, we read the 8-bytes-long
offsets one by one and pass them to the continuation, making sure
that we move the current position by 8 bytes before every recursive call.

We can readily use this function to inspect a top-level
constructor in a correct-by-construction manner like in
the following example.

\noindent
\begin{minipage}[t]{.4\textwidth}
  \ExecuteMetaData[SaferIndexed.idr.tex]{pureisleaf}
\end{minipage}\hfill\begin{minipage}[t]{.5\textwidth}
  \ExecuteMetaData[SaferIndexed.idr.tex]{isleaf}
\end{minipage}\medskip

Re-using the \IdrisFunction{Tree} description we
introduced in \Cref{fig:treedesc}, we defined a
pure \IdrisFunction{isLeaf} function checking that
an inductive tree is a leaf, together with its
effectful equivalent using \IdrisFunction{out}
to reveal the tree's top node thus allowing the
type-level call to the pure \IdrisFunction{isLeaf}
to reduce in each of the lambda-case's branches.
The use of \IdrisType{Singleton} guarantees that
we indeed return the appropriate boolean for the
tree we are pointing to.

\subsection{Offering a Convenient View}\label{sec:dataview}

We can combine \IdrisFunction{out} and \IdrisFunction{layer} to obtain
the \IdrisFunction{view} function we used in our introductory examples
in~\Cref{sec:seamless}.
This operation is not part of the trusted core: provided that
\IdrisFunction{poke} and \IdrisFunction{out} are correct then
it will automatically be correct-by-construction.
A (\IdrisType{View} \IdrisBound{cs} \IdrisBound{t}) value gives us
access to the (\IdrisType{Index} \IdrisBound{cs}) of
\IdrisBound{t}'s top constructor together with the corresponding
\IdrisFunction{Layer} of deserialised values and pointers to subtrees.

\ExecuteMetaData[SaferIndexed.idr.tex]{viewdata}

The implementation of \IdrisFunction{view} is unsurprising: we use
\IdrisFunction{out} to expose the top constructor index and a
\IdrisType{Pointer.Meaning} to the constructor's payload.
We then user \IdrisFunction{layer} to extract the full
\IdrisFunction{Layer} of deserialised values that the
pointer references.

\ExecuteMetaData[SaferIndexed.idr.tex]{view}

It is worth noting that although a \IdrisFunction{view} may be
convenient to consume, a performance-minded user may decide to
directly use the \IdrisFunction{out} and \IdrisFunction{poke}
combinators to avoid deserialising values that they do not need.
We present a case study in \Cref{appendix:rightmost} comparing the
access patterns of two implementations of the function fetching the
byte stored in a tree's rightmost node depending on whether we use
\IdrisFunction{view} or the lower level \IdrisFunction{poke} combinator.

By repeatedly calling \IdrisFunction{view}, we can define the
correct-by-construction generic deserialisation function: by
using \IdrisType{Singleton}, its type guarantees that we turn
a pointer to a tree \IdrisBound{t} into a runtime value known
to be equal to \IdrisBound{t}.

\ExecuteMetaData[SaferIndexed.idr.tex]{deserialise}

We can measure the benefits of our approach by comparing the runtime
of a function directly operating on buffers to its pure counterpart
composed with a deserialisation step.
For functions like \IdrisFunction{rightmost} that only explore a
very small part of the full tree, the gains are spectacular: the
process operating on buffers is exponentially faster than its
counterpart which needs to deserialise the entire tree first
(cf. \Cref{sec:timing}).

\subsection{Generic Fold}\label{sec:bufferfold}

The implementation of the generic \IdrisFunction{fold} over a tree stored
in a buffer is going to have the same structure as the generic fold over
inductive values: first match on the top constructor, then use \IdrisFunction{fmap}
to apply the fold to all the substructures and, finally, apply the algebra to
the result.
We start by implementing the buffer-based counterpart to \IdrisFunction{fmap}.
Let us go through the details of its type first.

\ExecuteMetaData[SaferIndexed.idr.tex]{fmaptype}

The first two arguments to \IdrisFunction{fmap} are similar to its pure
counterpart:
a description \IdrisBound{d}
and a (here runtime-irrelevant) function \IdrisBound{f}
to map over a \IdrisFunction{Meaning}.
Next we take a function which is the buffer-aware counterpart to \IdrisBound{f}:
given any runtime-irrelevant term \IdrisBound{t} and a pointer to it in a buffer,
it returns an \IdrisType{IO} process computing the value (\IdrisBound{f} \IdrisBound{t}).
Finally, we take a runtime-irrelevant meaning \IdrisBound{t}
as well as a pointer to its representation in a buffer and compute
an \IdrisType{IO} process which will return a value equal to
(\IdrisFunction{Data.fmap} \IdrisBound{d} \IdrisBound{f} \IdrisBound{t}).
We can now look at the definition of \IdrisFunction{fmap}.

\ExecuteMetaData[SaferIndexed.idr.tex]{fmapfun}

We poke the buffer to reveal the value the \IdrisType{Pointer.Meaning}
named \IdrisBound{ptr} is pointing at and then dispatch over the description
\IdrisBound{d} using the \IdrisFunction{go} auxiliary function.
If the description is \IdrisData{None} we match on the constructor
\IdrisData{I} of the \IdrisType{Poke'} which refines the type-level
term \IdrisBound{t} to the constructor \IdrisType{I} of the record
\IdrisType{True} and thus allows the \IdrisFunction{fmap} call to
reduce to \IdrisData{I}. We can therefore immediately return the
singleton wrapping the value \IdrisData{I}.
If the description is \IdrisData{Byte}, the value is left untouched
and so we can simply return it immediately.
If we have a \IdrisData{Prod} of two descriptions, we recursively apply
\IdrisFunction{fmap} to each of them and pair the results back.
Finally, if we have a \IdrisData{Rec} we apply the function operating
on buffers that we know performs the same computation as \IdrisBound{f}.

We can now combine \IdrisFunction{out} and \IdrisFunction{fmap} to compute
the correct-by-construction \IdrisFunction{fold}: provided an algebra for
a datatype \IdrisBound{cs} and a pointer to a tree \IdrisBound{t}
of type \IdrisBound{cs} stored in a buffer, we return an \IdrisType{IO}
process computing the same value as the pure \IdrisFunction{fold} would
have when applied to \IdrisBound{t}.

\ExecuteMetaData[SaferIndexed.idr.tex]{foldtype}

We first use \IdrisFunction{out} to reveal the constructor choice in the
tree's top node, we then recursively apply (\IdrisFunction{fold} \IdrisBound{alg})
to all the substructures by calling \IdrisFunction{fmap}, and we conclude by
applying the algebra to this result.

\ExecuteMetaData[SaferIndexed.idr.tex]{foldfun}

We once again (cf. \Cref{sec:genericfoldinductive}) had to
use \assertTotal{} because it is not obvious to
\idris{} that \IdrisFunction{fmap} only uses its argument on subterms.
This could have also been avoided by mutually defining \IdrisFunction{fold}
and a specialised version of
(\IdrisFunction{fmap} \IdrisKeyword{(}\IdrisFunction{fold} \IdrisBound{alg}\IdrisKeyword{)})
at the cost of code duplication and obfuscation.
We once again include such a definition in \Cref{appendix:safefold}.

\section{Serialising Data}\label{sec:serialising}

So far all of our example programs involved taking an inductive value
apart and computing a return value in the host language.
But we may instead want to compute another value in serialised form.
We include below one such example: a \IdrisFunction{map} function which
takes a function \IdrisBound{f} acting on bytes,
a pointer to a \IdrisFunction{Tree} named \IdrisBound{t} and returns a
serialisation process that will eventually produce another
\IdrisFunction{Tree} equal to the one obtained by applying that function
all of the bytes stored in \IdrisBound{t}'s nodes.

\ExecuteMetaData[SaferIndexed.idr.tex]{serialisedmap}\label{fig:serialised-map}

It calls the \IdrisFunction{view} we just defined to observe whether
the tree is a leaf or a node.
If it's a leaf, it returns a leaf.
If it's a node, it returns a node where the map has been recursively applied
to the left and right subtrees while the function \IdrisBound{f} has been applied
to the byte \IdrisBound{b}.


In this section we are going to spell out how we can define the high-level
constructs used above to allow users to write these correct-by-construction
serialisers.

\subsection{The Type of Serialisation Processes}

A serialisation process for a tree \IdrisBound{t} that belongs to the
datatype \IdrisBound{cs} is a function that takes a buffer
and a starting position and returns an \IdrisType{IO} process that
serialises the term in the buffer at that position and computes the
position of the first byte past the serialised tree.

\ExecuteMetaData[SaferIndexed.idr.tex]{serialising}

We do not expect users to define such processes by hand and in fact
prevent them from doing so by not exporting the \IdrisData{MkSerialising}
constructor.
Instead, we provide high-level, invariant-respecting combinators to safely
construct such serialisation processes.

\subsection{Building Serialisation Processes}\label{sec:serialnode}

Our main combinator is \IdrisFunction{(\#)}: by providing
a node's constructor index
and a way to serialise all of the node's subtrees,
we obtain a serialisation process for said node.
We will give a detailed explanation of \IdrisFunction{All} below.

\ExecuteMetaData[SaferIndexed.idr.tex]{serialisemu}

The keen reader may refer to the accompanying code to see the implementation.
Informally (cf. \Cref{sec:tree-serialisation} for the description of the format):
first we write the tag corresponding to the choice of constructor,
then we leave some space for the offsets,
in the meantime we write all of the constructor's arguments and collect the offsets
associated to each subtree while doing so,
and finally we fill in the space we had left blank with the offsets
we have thus collected.

The \IdrisFunction{All} quantifier performs the pointwise lifting of a predicate over
the functor described by a \IdrisType{Desc}. It is defined by induction over
the description.

\ExecuteMetaData[SaferIndexed.idr.tex]{allquant}

If the description is \IdrisData{Byte} we only demand that we have a runtime
copy of the byte so that we may write it inside a buffer. This is done using
the \IdrisType{Singleton} family discussed in \Cref{sec:datasingleton}.
If the description is \IdrisData{Rec} then we demand that the
predicate holds.
If the description is either \IdrisData{None} or \IdrisData{Prod} then we
use once again an auxiliary family purely for ergonomics.
It is defined mutually with \IdrisFunction{All} and does the expected structural
operation.

\ExecuteMetaData[SaferIndexed.idr.tex]{dataallquant}

It should now be clear that
(\IdrisFunction{All} \IdrisKeyword{(}\IdrisFunction{description} \IdrisBound{k}\IdrisKeyword{)}
(\IdrisType{Serialising} \IdrisBound{cs}\IdrisKeyword{)}) indeed corresponds
to having already defined a serialisation process for each subtree.

This very general combinator should be enough to define all the
serialisers we may ever want.
By repeatedly pattern-matching on the input tree and using \IdrisFunction{(\#)},
we can for instance define the correct-by-construction generic serialisation function.

\ExecuteMetaData[SaferIndexed.idr.tex]{serialise}

We nonetheless include other combinators purely for performance reasons.

\subsection{Copying Entire Trees}\label{sec:copy}

We introduce a \IdrisFunction{copy} combinator for trees that we want to
serialise as-is and have a pointer for.
Equipped with this combinator, we are able to easily write e.g.
the \IdrisFunction{swap} function which takes a binary tree apart
and swaps its left and right branches (if the tree is non-empty).

\ExecuteMetaData[SaferIndexed.idr.tex]{swap}

We could define this \IdrisFunction{copy} combinator at a high level
either by composing
\IdrisFunction{deserialise} and \IdrisFunction{serialise},
or by interleaving calls to \IdrisFunction{view} and \IdrisFunction{(\#)}.
This would however lead to a slow implementation that needs to
traverse the entire tree in order to simply copy it.

Instead, we implement \IdrisFunction{copy} by using the
\IdrisFunction{copyData} primitive for \IdrisType{Buffer}s
present in \idris{}'s standard library.
This primitive allows us to grab a slice of the source buffer
corresponding to the tree
and to copy the raw bytes directly into the target buffer.
This use of an unsafe primitive makes \IdrisFunction{copy} part
of the trusted core for this library.

\ExecuteMetaData[SaferIndexed.idr.tex]{copy}

This is the one combinator that crucially relies
on our format only using offsets and not absolute addresses
and on the accuracy of the size information we have been keeping
in \IdrisType{Pointer.Mu} and \IdrisType{Pointer.Meaning}.
As we can see in \Cref{sec:timing}, this is spectacularly faster than
a deep copying process traversing the tree.

\subsection{Executing a Serialisation Action}\label{sec:execserialising}

Now that we can describe actions serialising a value to a buffer,
the last basic building block we are still missing is a function actually
performing such actions.
This is provided by the \IdrisFunction{execSerialising} function
declared below.

\ExecuteMetaData[SaferIndexed.idr.tex]{execSerialising}

By executing a (\IdrisType{Serialising} \IdrisBound{cs} \IdrisBound{t}), we
obtain an \IdrisType{IO} process returning a pointer to the tree \IdrisBound{t}
stored in a buffer.
We can then either compute further with this tree (e.g. by calling
\IdrisFunction{sum} on it), or write it to a file for safekeeping
using the function \IdrisFunction{writeToFile}
introduced in \Cref{sec:writetofile}.

\subsection{Evaluation Order}

The careful reader may have noticed that we can and do run arbitrary \IdrisType{IO}
operations when building a value of type \IdrisType{Serialising}
(cf. the \IdrisFunction{map} example in \Cref{fig:serialised-map} where we perform
a call to \IdrisFunction{view} to inspect the input's shape).

This is possible thanks to \idris{} elaborating \IdrisKeyword{do}-blocks using
whichever appropriate bind operator is in scope. In particular, we have defined
the following one to use when building a serialisation process:

\ExecuteMetaData[SaferIndexed.idr.tex]{serialisingbind}

By using this bind we can temporarily pause writing to the buffer to make
arbitrary \IdrisType{IO} requests to the outside world.
In particular, this allows us to interleave reading from the original buffer
and writing into the target one thus having a much better memory footprint than
if we were to first use the \IdrisType{IO} monad to build in one go the whole
serialisation process for a given tree and then execute it.

\newcommand{\graphOf}[3]{\begin{tikzpicture}
  \begin{axis}
    [ xlabel = depth
    , ylabel = #3 (ns)
    , ymode = log
    , legend pos = north west
    ]
    \addplot table
      [ grid = major
      , x = size
      , y = #1
      , col sep = comma
      ] {#3.csv};
    \addplot table
      [ grid = major
      , x = size
      , y = #2
      , col sep = comma
      ] {#3.csv};
  \legend{#1, #2}
  \end{axis}
\end{tikzpicture}}

\newcommand{\graphOfThree}[4]{\begin{tikzpicture}
  \begin{axis}
    [ xlabel = depth
    , ylabel = #4 (ns)
    , ymode = log
    , legend pos = north west
    ]
    \addplot table
      [ grid = major
      , x = size
      , y = #1
      , col sep = comma
      ] {#4.csv};
    \addplot table
      [ grid = major
      , x = size
      , y = #2
      , col sep = comma
      ] {#4.csv};
    \addplot table
      [ grid = major
      , x = size
      , y = #3
      , col sep = comma
      ] {#4.csv};
  \legend{#1, #2, #3}
  \end{axis}
\end{tikzpicture}}

\section{Benchmarks}\label{sec:timing}

Now that we have the ability to read, write, and program directly
over trees stored in a buffer we can run some experiments to see
whether this allows us to gain anything over the purely functional
programming style.

For all of these tests we generate full trees for each depth,
labeling its nodes with the bytes 0, 1, etc. in a depth-first
left-to-right manner.

Each test is run 25 times in a row, and the duration averaged.
We manually run chezscheme's garbage collector before the start of
each time measurement.

All of our plots use a logarithmic y axis because the runtime of the
deserialisation-based functions is necessarily exponential in the depth
of the full tree.


\subsection{Pointer vs. Data}

We will first look at the runtime characteristics of various functions
implemented using the views defined in \Cref{sec:poking}.
We compare the time it takes to run the composition of deserialising
the tree and applying the pure function to the time it takes to run
its pointer-based counterpart.

\subsubsection{Traversing the Full Tree}

The \IdrisFunction{sum} function visits all of the tree's nodes to
add up all of the bytes that are stored.

\smallskip\noindent
\begin{minipage}{.5\textwidth}
  \ExecuteMetaData[SaferIndexed.idr.tex]{tsum}
\end{minipage}\hfill
\begin{minipage}{.45\textwidth}
  \graphOf{data}{pointer}{sum}
\end{minipage}\smallskip

Because this function explores the entirety of the tree, the
difference between the deserialisation-based and the pointer-based
functions is minimal.

\subsubsection{Skipping Most of the Tree}

The \IdrisFunction{rightmost} function goes down the tree's right hand
and returns the byte stored in its rightmost node.

\smallskip\noindent
\begin{minipage}{.5\textwidth}
  \ExecuteMetaData[SaferIndexed.idr.tex]{rightmost}
\end{minipage}\hfill
\begin{minipage}{.45\textwidth}
  \graphOf{data}{pointer}{rightmost}
\end{minipage}\smallskip

Because this function only traverses the rightmost branch of the tree,
the pointer based implementation is able to run in linear time by
efficiently skipping past every left subtree.
This is effectively an exponential speedup compared to the implementation
that first fully deserialises the tree.

\subsubsection{Exploring a Bounded Fragment of the Tree}


Let us for a change look at a function returning a richly typed result.
The \IdrisFunction{find} function looks for a given byte in a tree
and returns a path to it if it can.
The notion of \IdrisType{Path} is defined inductively: if the
\IdrisBound{tgt} byte we are looking for is in the node then
we can use \IdrisData{Here}; otherwise we can take a turn down
the left or right subtree using \IdrisData{TurnL} and \IdrisData{TurnR}
respectively and provide a path to the target byte in the corresponding
subtree.

\ExecuteMetaData[SaferIndexed.idr.tex]{path}

To save space we only present a semi-decision procedure but this could
be generalised to a full decision procedure.
Given that \IdrisType{Path} is a very informative type, in this test
case the version manipulating pointers does not bother using the
\IdrisType{Singleton} family.

\smallskip\noindent
\begin{minipage}{.5\textwidth}
  \ExecuteMetaData[SaferIndexed.idr.tex]{tfind}
\end{minipage}\hfill
\begin{minipage}{.45\textwidth}
  \graphOf{data}{pointer}{find}
\end{minipage}\smallskip

We run the test using the byte 120.
Due to the way our test trees are generated, we will only need to
explore at most 121 of the tree's nodes before finding this byte.
Unsurprisingly we observe that the pointer-based function is
constant time while the one operating over data is exponential
due to the deserialisation step.



\subsection{Serialising}

Let us now turn to the time characteristics of the serialisation primitives
defined in \Cref{sec:serialising}.
We compare the time it takes to run the composition of deserialising
the tree, applying the pure function, and serialising the result
to the time it takes to run its pointer-based counterpart using the
\IdrisType{Serialising} framework.

\subsubsection{Traversing the Full Tree}

The \IdrisFunction{map} function applies a byte-to-byte function
to all of the bytes stored in a tree's node.
We run it using the \IdrisKeyword{(}\IdrisFunction{+} \IdrisData{100}\IdrisKeyword{)}
function.

\noindent
\begin{minipage}{.5\textwidth}
  \ExecuteMetaData[SaferIndexed.idr.tex]{serialisedmapcompact}
\end{minipage}\hfill
\begin{minipage}{.45\textwidth}
  \graphOf{data}{pointer}{map}
\end{minipage}

We can see that both approaches yield a similar runtime: everything is
dominated by traversing the whole tree and building the resulting output.
This is where a framework allowing for destructive updates may help
write faster functions.
This also has untapped opportunities for parallelism.

\subsubsection{Using the Copy Primitive}

The \IdrisFunction{swap} function takes a tree and, if it is non-empty,
swaps its left and right subtrees.
For this test case we compare an implementation using the
\emph{primitive} \IdrisFunction{copy} operator we introduced in \Cref{sec:copy},
one using a \emph{pointer}-based copy that interleaves exposing the head
constructor and serialising it,
and a copy going via the \emph{data} representation by fully deserialising
the tree before re-serialising it.

\noindent
\begin{minipage}{.5\textwidth}
  \ExecuteMetaData[SaferIndexed.idr.tex]{swap}
\end{minipage}\hfill
\begin{minipage}{.45\textwidth}
  \graphOfThree{data}{pointer}{primitive}{swap}
\end{minipage}

We can see that, the \emph{pointer} and \emph{data} variants yield a similar
runtime as they need to traverse the full tree whereas the primitive
one based on copying raw buffer slices is significantly faster.

\section{Conclusion}

We have seen how, using a universe of descriptions indexed by their static
and dynamic sizes, we can define a precise language of values serialised in
a buffer.
This allowed us to develop a library to manipulate such trees in a seamless,
correct, and generic manner either using low-level combinators like
\IdrisFunction{poke} or high-level programs like a data-polymorphic
\IdrisFunction{fold}.
We then provided users with convenient tools to write serialisation processes
thus allowing them to compositionally build correct-by-construction values
stored in buffers.
Finally, we demonstrated that operating directly over serlialised data can
yield exponential speedups for functions that do not need to explore the
whole tree they receive as an input.

\subsection{Related Work}

This work sits at the intersection of many domains:
data-generic programming,
the efficient runtime representation of functional data,
programming over serialised values,
and the design of serialisation formats.
Correspondingly, a lot of related work is worth discussing.
In many cases the advantage of our approach is precisely that it is
at the intersection of all of these strands of research:
we do it generically, seamlessly, in a correct-by-construction manner,
and use actual buffers.

\subsubsection{Data-Generic Programming}

There is a long tradition of data-generic
programming~\citep{DBLP:conf/ssdgp/Gibbons06} and we will mostly focus here
on the approach based on the careful reification of a precise universe of
discourse as an inductive family in a host type theory,
and the definition of generic programs by induction over this family.

One early such instance is Pfeifer and Rue{\ss}'
`polytypic proof construction'~\citeyearpar{DBLP:conf/tphol/PfeiferR99}
meant to replace unsafe meta-programs deriving recursors
(be they built-in support, or user-written tactics).
Their approach requires the definition of an entirely new core language
to support polytypic function definitions while ours is completely
non-invasive.
It would be interesting to port some of their motivating examples to our
setting such as their generic data compression scheme for serialised
tree representations.

In his PhD thesis, Morris~\citeyearpar{DBLP:phd/ethos/Morris07} declares
various universes for strictly positive types and families and defines by
generic programming further types (the type of one-hole contexts),
modalities (the universal and existential predicate lifting over the functors
he considers), and functions (map, boolean equality).
We reuse some of these constructions already (e.g. the universal predicate
lifting over functors is the \IdrisType{All} type family we used to define
serialisation combinators in \Cref{sec:serialising}) and the other ones are
obvious candidates for future work.
The type of one-hole contexts is in particular extremely useful to implement
tail-recursive operations traversing a data structure.

L{\"{o}}h and Magalh{\~{a}}es~\citeyearpar{DBLP:conf/icfp/LohM11} define a more
expressive universe over indexed functors that is closed under composition
and fixpoints.
They also detail how to define additional generic construction such as a
proof of decidable equality, various recursion schemes, and zippers.
This work, quite similar to our own in its presentation, offers a natural
candidate universe for us to use to extend our library.

Recent efforts by Escot and Cockx~\citeyearpar{DBLP:journals/pacmpl/EscotC22}
have demonstrated that it is possible, through reflection, to offer
a neater user interface by transporting constructions between (a subset of)
the native Agda inductive types and their rich universe of descriptions.
This could potentially provide a more convenient way to program using
our library: instead of having to define the functions in the specification
layer over \IdrisType{Data.Mu} trees, users could program against bona fide
host language datatypes while still benefiting from the proven-correct
generic programs.

\subsubsection{Efficient Runtime Representation of Inductive Values}

Although not dealing explicitly with programming over serialised data,
Monnier's work~\citeyearpar{DBLP:conf/icfp/Monnier19} with its focus on performance and
in particular on the layout of inductive values at runtime,
partially motivated our endeavour.
Provided that we find a way to get the specialisation and partial evaluation
of the generically defined views, we ought to be able to achieve --purely in
user code-- Monnier's vision of a representation where $n$-ary tuples have
constant-time access to each of their component.

Baudon, Radanne, and Gonnord~\citeyearpar{ACM:conf/icfp/Baudon23} adopted
a radically different approach to ours: we provide users with a
uniform representation that requires no setup on their part,
while \textsc{Ribbit} instead gives power users absolute freedom
to define their own data layouts.
The project provides seamless integration: users annotate their
data declarations with the data layout they want.
The programming experience is left unchanged: users write nested
pattern-matching programs without ever needing to explicitly talk
about the layout.
They cover a universe of monomorphic immutable algebraic datatypes
fairly similar to ours, however the annotations language is expressive
enough to talk about e.g. principled struct packing, bit stealing,
unboxing, untagged unions, or word alignment.
Their approach is really impressive but requires modifying the
whole compiler pipeline down to producing LLVM IR while ours
lives purely in userland.

\textsc{Dargent}~\citep{DBLP:journals/pacmpl/ChenLOKMJR23} is a similar
high-level framework empowering users to dictate the data layout
that the generated C code shall use.
This means that, just like in our project, users do not need to
deserialise a value before being able to process it.
This work has the added benefit, compared to \textsc{Ribbit}, that
it automatically generates an Isabelle/HOL correctness proof that
the generated C code is a refinement of the original program.

Allais~\citeyearpar{DBLP:conf/esop/Allais23}
demonstrates how one can encode an
invariant-rich inductive type using builtin types
in \idris{} and recover high level programming principles
by using views.
They rely heavily on Quantitative Type Theory to precisely
control what parts of the encoding are to be erased during
compilation.
Their approach, dealing with a single inductive family in a
tutorial-style paper, is ad-hoc but complementary to ours
and shows how to use bit packing in a certified way.

\subsubsection{Working on Serialised Data}

Swierstra and van Geest~\citeyearpar{DBLP:conf/icfp/GeestS17}
define in \agda{} a rich universe of data descriptions
and generically implement serialising and deserialising,
proving that the latter is a left inverse to the former.
Their universe is powerful enough that later parts of a
description can be constrained using the \emph{value}
associated to ealier ones.
We will be able to rely on this work when extending our
current universe to one for type families.
In turn, our approach could be ported to their setting
to avoid the need to fully deserialise the data in order
to process it.

LoCal~\citep{DBLP:conf/pldi/VollmerKRS0N19} is the work that originally
motivated the design of this library.
We have demonstrated that generic programming within a dependently typed
language can yield the sort of benefits other language can only achieve
by inventing entirely new intermediate languages and compilation schemes.

LoCal was improved upon with a re-thinking of the serialisation scheme
making the approach compatible with parallel
programming~\citep{DBLP:journals/pacmpl/KoparkarRVKN21}.
This impressive improvement is a natural candidate for future work on our
part: the authors demonstrate it is possible to reap the benefits of
both programming over serialised data
and dividing up the work over multiple processors
with almost no additional cost in the case of a purely sequential execution.

\subsubsection{Serialisation Formats}

The PADS project~\citep{DBLP:conf/popl/MandelbaumFWFG07} aims to let users
quickly, correctly, and compositionally describe existing formats they
have no control over.
As they remind us, ad-hoc serialisation formats abound be it in
networking, logging, billing, or as output of measurement equipments
in e.g. gene sequencing or molecular biology.
Our current project is not offering this kind of versatility as we have
decided to focus on a specific serialisation format with strong
guarantees about the efficient access to subtrees.
But our approach to defining correct-by-construction components could
be leveraged in that setting too and bring users strong guarantees about
the traversals they write.

ASN.1~\citep{MANUAL:book/larmouth1999} gives users the ability to define
a high-level specification of the exchange format (the `abstract syntax')
to be used in communications without the need to concern themselves with
the actual encoding as bit patterns (the `transfer syntax').
This separation between specification and implementation means that parsing
and encoding can be defined once and for all by generic programming
(here, a compiler turning specifications into code in the user's host
language of choice).
The main difference is once again our ability to program in a
correct-by-construction manner over the values thus represented.

Yallop's automatic derivation of serialisers using an OCaml
preprocessor~\citep{DBLP:conf/ml/Yallop07} highlights the importance
of empowering domain experts to take advantage of the specifics of
the problem they are solving to minimise the size of the encoded data.
By detecting sharing using a custom equality function respecting
$\alpha$-equivalence instead of the default one, he was able to
serialise large lambda terms using only
a quarter of the bytes required
by the OCaml's standard library marshaller.

\subsubsection{Type Theory for Data Layout}

Petersen, Robert, Crary, and Pfenning~\citeyearpar{DBLP:conf/popl/PetersenHCP03}
design an elegant type theory based on ordered linear logic
to describe the memory layout and an effectful functional
language in the style of Moggy's
metalanguage~\citeyearpar{DBLP:journals/iandc/Moggi91}.
This empowers users to carefully control their data's memory
layout.
This logic-based approach supports the explicit reservation,
initialisation, and allocation of new memory locations all
without needing to explicitly talk about an underlying
heap-based memory model.
Their language's intended semantics shares our economic flat
structure but only for non-recursive data: components of a
pair are stored contiguously, and values of the unit type are
absent at runtime.
In their extended technical report~\citeyearpar{MANUAL:report/cmu/PetersenHCP03}
they describe support for sum types, both tagged and untagged
(through an internalised notion of distinguishable types).
They also describe a simple scheme to support iso-recursive types
by giving up on a flat memory representation and insisting that
every node is behind a pointer indirection.
Their appproach requires a dedicated language implementation
while our definitions live purely in client code, defining
an embedded domain specific language as a library.
They support destructive memory updates but do not offer any
support for internally proving the properties of the programs
one defines.

\subsection{Limitations and Future Work}

Although our design is already proven to be functional by two implementations
in \idris{} and \agda{} respectively, we can always do better.
In this section we are going to see what benefits future work could bring
across the whole project.

\subsubsection{A Smaller Trusted Core}

Our current trusted core amounts to the implementations of
\IdrisFunction{poke} (defined in \Cref{sec:poke}),
\IdrisFunction{out} (defined in \Cref{sec:out}),
the serialisation combinator \IdrisFunction{(\#)} (defined in \Cref{sec:serialnode}),
\IdrisFunction{copy} (defined in \Cref{sec:copy}),
and \IdrisFunction{execSerialising} (defined in \Cref{sec:execserialising}).

As observed in \cref{sec:pointers} the indices used in
\IdrisType{Pointer.Mu} and \IdrisType{Pointer.Meaning}
are phantom indices~\citep{DBLP:conf/dsl/LeijenM99}.
This is the reason why our arithmetic operations on pointers are
unverified and we have to resort to escape hatches to deploy the
observations made by reading bytes in the buffer in order to
refine indices.

A further engineering effort could allow us to move this trusted
core slightly down the stack by defining a type of buffers indexed
by a (phantom) list of bytes corresponding to their content together
with trusted reading primitives whose behaviour is expressed
in terms of that phantom list.
Reusing existing work on correct-by-construction
serialisers~\citep{DBLP:conf/icfp/GeestS17} we could
then prove that the implementations of our basic building blocks are
indeed safe provided that the buffer primitives are correct.

\subsubsection{A More Robust Library}\label{sec:limitation-robust}

For sake of ease of presentation we have not dealt with issues necessitating
buffer resizing: in \Cref{sec:serialising},
we defined \IdrisFunction{execSerialising}
by allocating a fixed size buffer and not worrying whether the whole content
would fit.
A real library would need to adopt a more robust approach
akin to the one used in the implementation of \idris{}'s own serialisation
code: whenever we are about to write a byte to the buffer, we make sure there
is either enough space left or we grow it.
Note that our Agda port does not suffer from this limitation as it
can rely on Haskell's {\usestt\texttt{bytestring}}
library~\citep{DBLP:conf/padl/CouttsSL07} and use its
{\usestt\texttt{Builder}} type.

In our library, the data types descriptions currently need to be defined
as values in the host language.
This opens up the opportunity for bugs if, say, we write a server in
\idris{} and a client in \agda{} and accidentally use two slightly
different descriptions in the projects.
This could be solved at the language level by equipping our dependently typed
languages with dependent type providers like Idris~1
had~\citep{DBLP:conf/icfp/Christiansen13}.
This way the format could be loaded at compile time from the same file thus
ensuring all the components are referring to the exact same specification.

\subsubsection{A More Efficient Library}

Looking at the code generated by \idris{}, we notice that our generic programs
are not specialised and partially evaluated even when the types they are working
on are statically known.
Refactoring the library to use a continuation-passing-style approach does help
the compiler generate slightly more specialised code but the results are in our
opinion not good enough to justify forcing users to program in this more
cumbersome style.
A possible alternative would be to present users with macros rather than
generic programs so that the partial evaluation would be guaranteed to
happen at typechecking time. This however makes the process of defining
the generic programs much more error prone.
A more principled approach would be to extend \idris{} with a proper
treatment of staging e.g. by using a two-level type theory as suggested
by Kov{\'{a}}cs~\citeyearpar{DBLP:journals/pacmpl/Kovacs22}.

Our serialisation format has been designed to avoid pointer-chasing and
thus ensures entire subtrees can be easily copied by using the raw bytes.
Correspondingly it currently does not support sharing.
This could however be a crucial feature for trees with a lot of duplicated
nodes and we would like to allow users to, using the same interface,
easily pick between different serialisation formats so that the library
ends up using the one that suits their application best.
To this end, we could take inspiration from Yallop's
definition of preprocessors generating serialisers~\citep{DBLP:conf/ml/Yallop07}.
It maintains an object map containing the already serialised nodes and uses it
to maximally detect sharing and maintain it both when serialising and deserialising.

Our current approach allows us to define a correct-by-construction
\IdrisFunction{sum} operating directly on serialised data but it
does not eliminate the call stack used in the naïve functional
implementation.
Converting a fold to a tail recursive function in a generic manner
is a well studied problem and the existing
solutions~\citep{DBLP:conf/popl/McBride08,DBLP:conf/icfp/CortinasS18}
should be fairly straightforward, if time-consuming, to port to our setting.

\subsubsection{A More Expressive Universe of Descriptions}

We have used a minimal universe to demonstrate our approach but a practical
application would require the ability to store more than just raw bytes.
An easy extension is to add support for all of the numeric types of
known size that \idris{} offers
(\IdrisType{Bits\{8,16,32,64\}}, \IdrisType{Int\{8,16,32,64\}}),
for \IdrisType{Bool}
as well as a unbounded data such as \IdrisType{Nat}, or \IdrisType{String}
as long as an extra offset is provided for each value.

The storage of values smaller than a byte (here \IdrisType{Bool}) naturally
raises the question of bit packing: why store eight booleans as eight bytes
when they could fit in a single one?
\iftoggle{BLIND}{Allais'}{Our} recent work~\citep{DBLP:conf/esop/Allais23}
on the efficient runtime representation of inductive families as values
of \idris{}'s primitive types points us in the direction of a solution.

A natural next candidate is a universe allowing the definition of parametrised
types~\citep{DBLP:conf/icfp/LohM11}: we should be able to implement functions over arbitrary
(\IdrisType{List} \IdrisBound{a}) values stored in a buffer,
provided that we know that \IdrisBound{a} is serialisable.
This was already an explicit need in ASN.1~\citep{MANUAL:book/larmouth1999},
reflecting that protocols often leave `holes' where the content of
the protocol's higher layer is to be inserted.

Next, we will want to consider a universe of indexed data:
we can currently natively model algebraic datatypes such as lists or trees,
and we can use the host language to compute the description
of vectors by induction on their length,
but we cannot model arbitrary type families~\citep{DBLP:journals/fac/Dybjer94}
e.g. correct-by-construction red-black trees.

Last but not least we may want to have a universe of descriptions closed
under least fixpoints~\citep{DBLP:phd/ethos/Morris07}
in order to represent rose trees for instance.

\subsubsection{A More Expressive Library}

Using McBride's generalisation of one hole contexts~\citep{DBLP:conf/popl/McBride08}
we ought to be able to give a more precise type to the combinator
\IdrisFunction{(\#)} used to build serialisation processes.
When defining the serialisation of a given subtree, we ought to have access to
pointers to the result of serialising any subtree to the left of it. In particular
this would make building complete binary trees a lot faster by allowing us to rely
on \IdrisFunction{copyData} for duplicating branches rather than running the computation
twice.

Last, but not least we currently do not support in-place updates to
the data stored in a buffer.
This could however be beneficial for functions like \IdrisFunction{map}.
It remains to be seen whether we can somehow leverage \idris{}'s
linear quantity annotation to provide users with serialised value that
can be safely updated in place.
This would turn our ongoing metaphor involving
Hoare triples~\citep{DBLP:journals/cacm/Hoare69},
heap pointers,
and separation logic~\citep{DBLP:conf/lics/Reynolds02}
into a bona fide shallow embedding.
Poulsen, Rouvoet, Tolmach, Krebbers, and Visser's
pioneering work~\citep{DBLP:journals/pacmpl/PoulsenRTKV18,DBLP:phd/basesearch/Rouvoet21}
on definitional interpreters for languages with references
and the use of a shallowly embedded separation logic to
minimise bookkeeping give us a clear set of techniques
to adapt to our setting.

\section*{Acknowledgements}
\iftoggle{BLIND}{remarks from friends}{
  We would like to thank Wouter Swierstra, Fredrik Nordvall-Forsberg,
  as well as the anonymous reviewers for their helpful comments on
  various drafts of this paper.
}

\iftoggle{BLIND}{funding and data}{
This research was partially funded by the Engineering and Physical Sciences
Research Council (grant number EP/T007265/1).
}

\bibliographystyle{JFPlike}
\bibliography{paper}

\appendix

\section{Safe Implementations of fold}
\label{appendix:safefold}

We include below the alternative definitions of \IdrisFunction{fold}
(respectively processing inductive data and data stored in a buffer)
which are seen as total by \idris{}.
Each of them is mutually defined with what is essentially
the supercompilation of
(\IdrisKeyword{\textbackslash} \IdrisBound{d} \IdrisKeyword{=>} \IdrisFunction{fmap} \IdrisBound{d} (\IdrisFunction{fold} \IdrisBound{alg})).

\ExecuteMetaData[SafeFolds.idr.tex]{datafold}

\ExecuteMetaData[SafeFolds.idr.tex]{pointerfold}

\section{Access Patterns: Viewing vs. Poking}\label{appendix:rightmost}

In this example we implement \IdrisFunction{rightmost}, the function walking
down the rightmost branch of our type of binary trees and returning the
content of its rightmost node (if it exists).

The first implementation is the most straightforward: use \IdrisFunction{view}
to obtain the top constructor as well as an entire layer of deserialised values
and pointers to substructures and inspect the constructor.
If we have a leaf then there is no byte to return.
If we have a node then call \IdrisFunction{rightmost} recursively and inspect the
result: if we got \IdrisData{Nothing} back we are at the rightmost node and can
return the current byte, otherwise simply propagate the result.

\ExecuteMetaData[SaferIndexed.idr.tex]{viewrightmost}

In the alternative implementation we use \IdrisFunction{out} to expose the top
constructor and then, in the node case, call \IdrisFunction{poke} multiple times
to get our hands on the pointer to the right subtree.
We inspect the result of recursively calling \IdrisFunction{rightmost} on this
subtree and only deserialise the byte contained in the current node if the result
we get back is \IdrisData{Nothing}.

\ExecuteMetaData[SaferIndexed.idr.tex]{pokerightmost}

This will give rise to two different access patterns: the first function will
have deserialised all of the bytes stored in the nodes along the tree's
rightmost path whereas the second will only have deserialised the rightmost byte.
Admittedly deserialising a byte is not extremely expensive but in a more realistic
example we could have for instance been storing arbitrarily large values in
these nodes. In that case it may be worth trading convenience for making sure we
are not doing any unnecessary work.

\label{lastpage01}

\end{document}